\documentclass[12pt]{article}

\usepackage{scicite,graphicx}
\usepackage{siunitx}
\usepackage{times}

\newenvironment{sciabstract}{%
\begin{quote} }
{\end{quote}}

\title{An excess of small-scale gravitational lenses observed in galaxy clusters}

\author{Massimo Meneghetti,$^{1\ast,2,3}$ Guido Davoli$^{1,4}$, Pietro Bergamini$^1$, Piero Rosati$^{5,1}$,\\
Priyamvada Natarajan$^{6}$, Carlo Giocoli$^{1,2,7}$, Gabriel B. Caminha$^8$, \\ 
R. Benton Metcalf$^7$, Elena Rasia$^{9,10}$, Stefano Borgani$^{9,10,11,12}$,  \\ 
Francesco Calura$^1$, Claudio Grillo$^{13,14}$, Amata Mercurio$^{15}$, Eros Vanzella$^1$\\
\\
\footnotesize{$^{1}$Osservatorio di Astrofisica e Scienza dello Spazio di Bologna, Istituto Nazionale di Astrofisica}\\
\footnotesize{Via Gobetti 93/3, I-40129, Bologna, Italy}\\
\footnotesize{$^{2}$National Institute for Nuclear Physics, viale Berti Pichat 6/2, I-40127 Bologna, Italy}\\
\footnotesize{$^{3}$Division of Physics, Mathematics, \& Astronomy, California Institute of Technology, Pasadena, CA 91125, USA}\\
\footnotesize{$^{4}$Centro Euro-Mediterraneo sui Cambiamenti Climatici (CMCC), viale Berti Pichat 6/2, I-40127 Bologna, Italy}\\
\footnotesize{$^{5}$Dipartimento di Fisica e Scienza della Terra, Università di Ferrara, via Saragat 1, I-44122 Ferrara, Italy}\\
\footnotesize{$^{6}$Department of Astronomy, 52 Hillhouse Avenue, Steinbach Hall, Yale University, New Haven, CT 06511, USA}\\
\footnotesize{$^{7}$Dipartimento di Fisica e Astronomia, Università di Bologna, via Gobetti 93/2, 40129 Bologna, Italy}\\
\footnotesize{$^{8}$Kapteyn Astronomical Institute, University of Groningen, Postbus 800, 9700 AV Groningen, The Netherlands}\\
\footnotesize{$^{9}$Osservatorio Astronomico di Trieste, Istituto Nazionale di Astrofisica, Via Tiepolo, 11, I-34131 Trieste, Italy}\\
\footnotesize{$^{10}$Institute for Fundamental, Physics of the Universe, Via Beirut 2, 34014 Trieste, Italy}\\
\footnotesize{$^{11}$Department of Physics, University of Trieste, via Tiepolo 11, I-34131 Trieste, Italy}\\
\footnotesize{$^{12}$National Institute for Nuclear Physics, Via Valerio 2, I-34127 Trieste, Italy}\\
\footnotesize{$^{13}$Dipartimento di Fisica, Università  degli Studi di Milano, via Celoria 16, I-20133 Milano, Italy}\\
\footnotesize{$^{14}$Niels Bohr Institute, University of Copenhagen, Lyngbyvej 2, 4. sal 2100 Copenhagen, Denmark}\\
\footnotesize{$^{15}$Osservatorio Astronomico di Capodimonte, Istituto Nazionale di Astrofisica, Salita Moiariello, 16, I-80131 Napoli, Italy}\\
\\
\footnotesize{$^\ast$E-mail:  massimo.meneghetti@inaf.it}
}

\date{}

\newcommand{\lesssim}{\raisebox{-0.13cm}{~\shortstack{$<$ \\[-0.07cm]$\sim$}}~}

\begin{document} 

%\si{\angstrom}
%\SI{1}{\angstrom}

% Double-space the manuscript.

\baselineskip24pt

% Make the title.

\maketitle

% Place your abstract within the special {sciabstract} environment.

\begin{sciabstract}
Cold dark matter (CDM) constitutes most of the matter in the Universe. The interplay between dark and luminous matter in dense cosmic environments like galaxy clusters is studied theoretically using cosmological simulations. Observed gravitational lensing is used to test and characterize the properties of substructures - the small-scale distribution of dark matter - in clusters. An apt metric, the probability of strong lensing events produced by dark matter substructure, is devised and computed for 11 galaxy clusters.  We report that observed cluster substructures are more efficient lenses than predicted by CDM simulations, by more than an order of magnitude. We suggest that hitherto undiagnosed systematic issues with simulations or incorrect assumptions about the properties of dark matter could explain our results.
\end{sciabstract}

\section*{Main text:}
In the standard cosmological model, the matter content of the Universe is dominated by cold dark matter (CDM), collisionless particles that interact with ordinary matter (baryons) only through gravity. Gravitationally bound dark-matter halos form hierarchically with the most massive systems forming through mergers of smaller ones. As structure assembles in this fashion, large dark matter halos contain smaller-scale substructure in the form of embedded sub-halos. 

The most massive dark matter halos at the present time are galaxy clusters, with masses of $\sim 10^{14-15}$ M$_\odot$, where M$_\odot$ is in units of the mass of the sun which corresponds to $\sim 2\times10^{30}$ kg. Galaxy clusters contain about a thousand or so member galaxies that are hosted in sub-halos. The detailed spatial distribution of dark matter in galaxy clusters can be mapped through their observed gravitational lensing of distant background galaxies. When distant background galaxies are in near perfect alignment with the massive foreground cluster, strong gravitational lensing occurs. Strong lensing - non-linear effects produced by the deflection of light - results in multiple distorted images of individual background galaxies that are now routinely detected with Hubble Space Telescope (HST) imaging.

The probability and strength of these non-linear strong lensing effects can be calculated theoretically from simulations \cite{2017MNRAS.472.3177M} of structure formation. We test the predictions from simulations with observations of galaxy clusters combining lensing data from the HST with spectroscopic data from the Very-Large-Telescope (VLT).
Our observed sample of lensing clusters is split into three sets for this analysis: (i) a reference sample comprising three clusters with well-constrained mass distributions (mass models) - MACS J1206.2-0847 at redshift $z=0.439$ (MACSJ1206); MACS J0416.1-2403 at $z=0.397$ (MACSJ0416) and Abell S1063 at $z=0.348$ (AS1063)  \cite{2017arXiv170700690C,2016A&A...587A..80C,2017A&A...600A..90C,2017ApJ...842..132B,2019A&A...631A.130B}; (ii) a sample that includes the publicly available mass models for four Hubble Frontier Fields clusters [HFF, \cite{2017ApJ...837...97L}], namely Abell 2744 ($z=0.308$),  Abell 370 ($z=0.375$), MACS J1149.5+2223 (MACSJ1149, $z=0.542$), and MACS J0717.5+3745 (MACSJ0717, $z=0.545$); and (iii) four clusters from the Cluster Lensing and Supernova Survey with Hubble [CLASH, \cite{2012ApJS..199...25P}], with recent mass reconstructions by \cite{2019arXiv190305103C} (their “Gold” sample): RX J2129.7+0005 (RXJ2129, $z=0.234$), MACS J1931.8-2635 (MACSJ1931, $z=0.352$), MACS J0329.7-0211 (MACSJ0329, $z=0.450$), and MACS J2129.4-0741 (MACSJ2129, $z=0.587$). A color-composite image of MACSJ1206, one of the clusters in our reference sample (i), is shown in Fig. 1A. The images of the other clusters are shown in Figs. S1-S3.

Owing to their large masses, all these galaxy clusters act as powerful strong lenses producing an abundance of multiple images. To reconstruct their mass distributions, in addition to these images, extensive spectroscopic data are also available \cite{2016A&A...587A..80C,2015ApJ...812..114T}.
For each cluster lens, the membership of hundreds of galaxies in the cluster is confirmed spectroscopically and their redshifts have been measured, in addition to the identification of tens of multiply imaged background sources. 

Mass models for the reference cluster sample were constructed by using the publicly available parametric lens inversion code \textsc{Lenstool} \cite{2007NJPh....9..447J} and are presented in \cite{2019A&A...631A.130B}.
Clusters are modeled as a superposition of large-scale components to account for the large-scale cluster dark matter halos, and small-scale components that describe the substructure. We associate the spatial positions of cluster member galaxies with the locations of dark matter substructure. Additionally, the detailed mass distribution in these cluster galaxies is constrained by using stellar kinematics measurements from VLT spectroscopy.

The mass models for the clusters in the other two samples are built similarly \cite{methods}; however the mass distribution in the cluster member galaxies in these, unlike for the reference sample, is not constrained by using data from stellar kinematics. For the HFF sample, a suite of lensing mass models constructed independently by several groups are publicly available from the Mikulski Archive for Space Telescopes (MAST).  However, we used only those built using \textsc{Lenstool} for consistency in our analysis [e.g. \cite{2014ApJ...797...48J,2017MNRAS.468.1962N}]. For the “Gold” sample, we use the models published by \cite{2019arXiv190305103C}, also built with \textsc{Lenstool}.

The multiple images of distant sources lensed by foreground galaxy clusters have angular separations of the order of several tens of arc sec. The most dramatically distorted gravitational arcs occur near lines that enclose the inner regions of the cluster, referred to as critical lines. The size of the critical lines depends on the redshifts of the sources and delineate the region wherein strong lensing occurs. Substructures within each cluster act as smaller scale gravitational lenses embedded within the larger lens. If these substructures are massive enough and compact enough, they can also produce additional local strong lensing events on much smaller scales with separations of less than a few arc sec. These small-scale features are expected to appear around the critical lines produced by individual cluster galaxies. We refer to these localized features as Galaxy-Galaxy Strong Lensing (GGSL) events. Sufficiently high-resolution mass reconstructions are necessary to recover these smaller scale critical lines. For instance, Fig. 1A shows the detailed network of critical lines in MACSJ1206 for two possible source redshifts ($z=1$ and $z =7$ shown as solid and dashed curves respectively). The cluster produces a large-scale critical line extending to 15-30 arc sec and many smaller scale critical lines around individual substructures shown in the insets. The presence of secondary critical lines indicates that the substructures are centrally concentrated and massive enough to act as individual strong lenses. 

We identify three GGSL events in the core of the cluster MACSJ1206, shown in Fig. 1 B to D: a ring-shape image, called an Einstein ring, originating from a source at $z=1.42$; a triply imaged galaxy at $z=3.75$ \cite{2015ApJ...800...38G}; and an Einstein cross with four distinct images of a source at $z=4.99$. 
The consistency between the shapes of the GGSL events and the predicted critical lines from the lens modeling, both shown in Fig. 1 B to D, validates the assumption of our multi-scale mass model.

Just as the observed gravitational arcs are lensed images of distant galaxies, the critical lines are the lensed counterparts of the caustic lines \cite{2017MNRAS.472.3177M}, shown in Figs. 2B and D.  The caustics enclose the regions in which sources have to be located to be strongly lensed by substructures. We quantify the probability of observing GGSL events using the fraction of the area of the sky inside the caustics produced by substructures. 
Fig. 3 shows how the GGSL probability varies as a function of the source redshift for all clusters in our three samples. For MACSJ1206 (upper limit of the reference sample), it is $\sim 10^{-3}$ at $z>2$. This probability can in turn be converted into an expected number of GGSL events by making assumptions about the properties of the background source population of galaxies that can be lensed. Using  galaxies seen in the Hubble Ultra-Deep Field (HUDF) \cite{2015AJ....150...31R} as a representative template for the properties of the background lensed sources, we calculate that $\lesssim 3$ GGSL events should occur in MACSJ1206, in agreement with the observations. Equivalent estimates for MACSJ0416 and AS1063 predict $\sim 1$ and $\sim 0.9$ events, respectively. In these two cases, our calculations under-predict the number of observed GGSL events, as three candidate events have been reported in each of two clusters \cite{2017ApJ...842...47V,2018MNRAS.479.2630D}. 
This under-estimate likely arises as the HUDF may not be an appropriate template for background sources in these two clusters \cite{methods}. Nevertheless, we find that GGSL events are detected in multiple clusters. Twenty-four GGSL candidate events have been found in other CLASH clusters, including four events in MACSJ1149 and one event in each of the clusters MACSJ0717, RXJ2129, and MACSJ0329 \cite{2018MNRAS.479.2630D}. 

We next consider whether the observed number of GGSL events are consistent with theoretical predictions within the concordance cosmological model. We performed the same analysis and computed the GGSL probability for 25 simulated galaxy clusters, which have masses, redshifts, morphologies, and mass concentrations similar to those in our three observed samples \cite{methods}. The cosmological hydrodynamical simulations from which these simulated clusters are drawn \cite{2014MNRAS.438..195P} incorporate gas cooling, star-formation and energy feedback from supernovae and accreting super-massive black holes (SMBHs) - all standard ingredients at present.

Fig. 2 shows a comparison between the critical lines and the caustics of MACSJ1206 (panels A and B) and those of a simulated cluster of similar mass and concentration (panels C and D). MACSJ1206 has many more secondary critical lines within the observed area. The fractional area of the source plane that is enclosed by substructure caustics is noticeably larger in observed clusters compared to that predicted for the simulated sample, as is the probability of GGSL events. Fig. 3 shows that the GGSL probability differs by more than an order of magnitude between the observations and simulations.

We performed several tests to investigate potential sources for this discrepancy \cite{methods}. 
The results remain unchanged even when one key ingredient - energy feedback from active-galactic-nuclei powered by SMBH accretion - which alters the internal structure of halos is disabled in the simulations. This feedback suppresses star formation in substructures, altering the slope of their inner density profiles, making them less centrally concentrated and, hence, weaker gravitational lenses. Even without feedback, we are unable to bridge the gap between simulations and observations completely. In addition, simulations without feedback would be grossly discrepant from observations for other well measured quantities like the total fraction of baryons in clusters converted into stars. The mass and spatial resolutions of our simulations are sufficiently high to resolve the typical substructures included in the lensing mass models. We also exclude the possibility that the computed GGSL probability could be enhanced by unassociated halos along the line-of-sight (LOS) to these clusters. Including multiple lens planes in the models generated using cosmological simulations, we find that the substructure critical lines and caustics are negligibly affected by halos along the LOS. The observationally constrained lens models reproduce the shapes and sizes of the observed GGSL events. For instance, the model predicted image positions match within $\sim 0.5$ arc sec with what is seen. 

The discrepancy between observations and simulations may be due to issues with either the CDM paradigm or simulation methods. Gravitational lensing has previously been used to probe detailed properties of dark-matter halos associated with individual cluster galaxies [e.g. \cite{1997MNRAS.287..833N,NA98.1}]. Simulations show that the mass and radial distributions of sub-halos are nearly universal \cite{2004MNRAS.355..819G}. 
Varying results have been reported for the level of agreement between lens model predictions and simulations for other derived quantities, e.g. the mass distribution functions of substructure derived from lensing data agree with simulations but their radial distributions are more centrally concentrated in observed clusters than seen in simulations \cite{2015ApJ...800...38G,2017ApJ...842..132B,2017MNRAS.468.1962N}. Strong lensing clusters also contain more high-circular-velocity sub-halos (i.e. sub-halos with maximum circular velocities $V_{circ}>100$ km s$^{-1}$) compared with simulations \cite{2015ApJ...800...38G,2016ApJ...827L...5M,2017ApJ...842..132B}. The maximum circular velocity is given by 
\begin{equation}
    V_{circ}=\mathrm{max}\sqrt{\frac{GM(r)}{r}} \;,
\end{equation}
where $G$ is the gravitational constant, $M(r)$ is the galaxy mass profile and $r$ is the distance from the galaxy center. Fig. 4 shows that, in our lens models, observed galaxies have larger circular velocities than their simulated analogs at a fixed mass. This implies that dark matter sub-halos associated with observed galaxies are more compact than theoretically expected. Observed substructures also appear to be in closer proximity to the larger scale cluster critical lines. 
Explaining this difference requires the existence of a larger number of compact substructures in the inner regions of simulated clusters. 
Baryons and dark matter are expected to couple in the dense inner regions of sub-halos, leading to alterations in the small-scale density profile of dark matter, it could be that our current understanding of this interplay is incorrect. Alternatively, the difference could arise from incorrect assumptions about the nature of dark matter.

Previous discrepancies between the predictions of the standard cosmological model and data on small scales have arisen from observations of dwarf galaxies and of satellites of the Milky Way, namely the so called “missing satellite” \cite{1999ApJ...524L..19M,1999ApJ...522...82K}, “cusp-core” \cite{FL94.1}, “too-big-to-fail” problems \cite{2011MNRAS.415L..40B,2010MNRAS.406..896B}, and planes of satellite galaxies \cite{2018Sci...359..534M}. The discrepancy that we report is unrelated to these other issues. Previous studies revealed that observed small satellite galaxies were fewer in number and were less compact than expected from simulations; here, we find the opposite for cluster substructures. The GGSL events that we observe show that subhalos are more centrally concentrated than predicted by simulations i.e. there is an excess not a deficit. Hypotheses advocated to solve previous controversies on dwarf galaxy scales would only serve to exacerbate the discrepancy in GGSL event numbers that we report here.

Our results therefore require alternative explanations. One possibility is numerical effects arising from the resolution limits of simulations \cite{2018MNRAS.474.3043V}.
However, currently known numerical artefacts are not effective enough at disrupting satellites. We investigated this issue thoroughly \cite{methods} and found that it can change the predicted GGSL event rate at most by a factor of 2, which is insufficient to explain the nearly order of magnitude discrepancy that we find. These numerical artefacts would also appear on galactic scales, where they would in turn worsen the “missing satellite problem”. 

\begin{figure}
    \centering
    \includegraphics[width=0.9\textwidth]{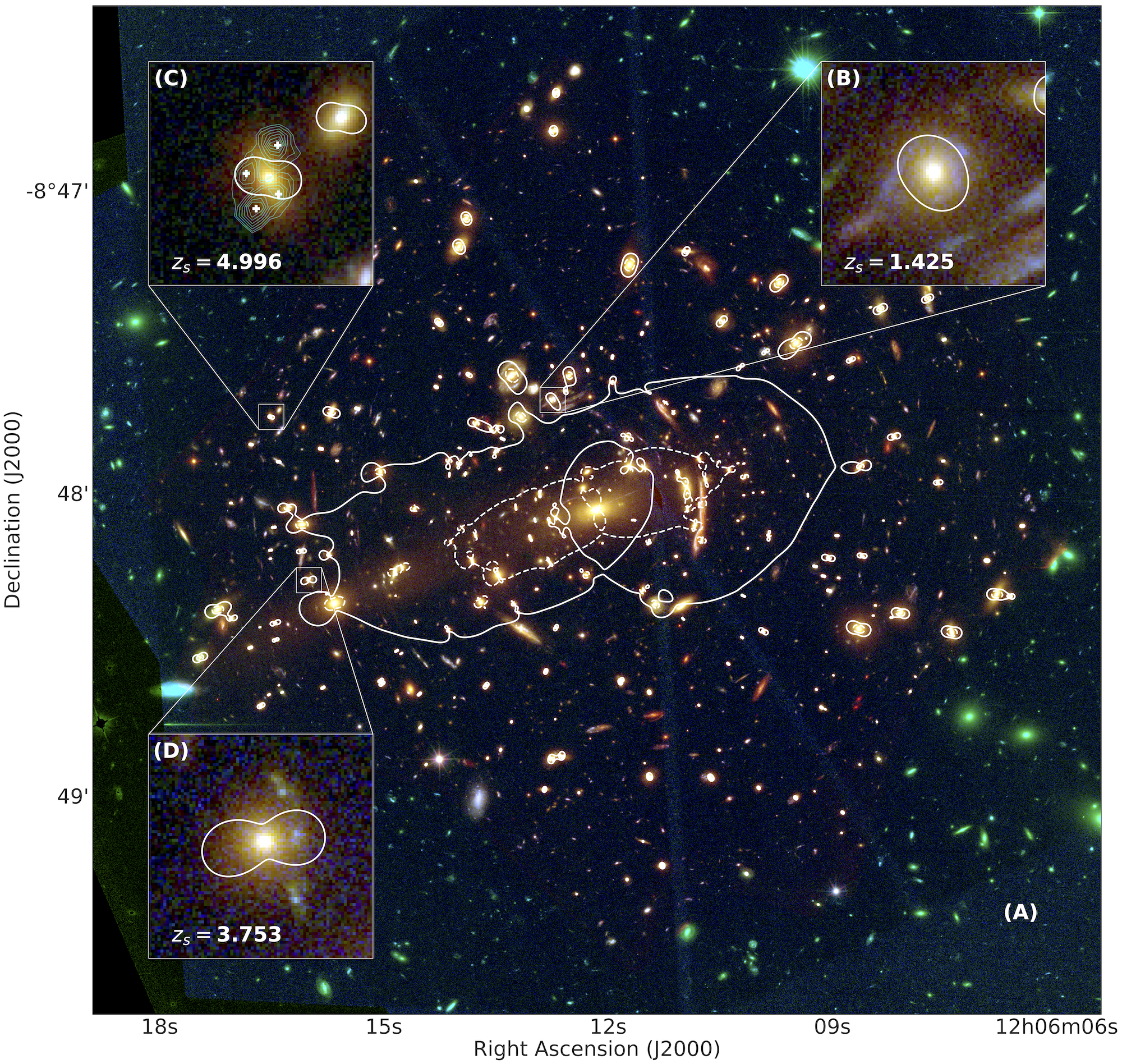}
    \caption{{\bf Color-composite image of the central region of the galaxy cluster MACSJ1206. (A to D)} The
image combines HST observations in the filters F105W, F110W, F125W, F140W, F160W (red channel), F606W,
F625W, F775W, F814W, F850LP (green channel), and F435W and F475W (blue channel). The dashed and
solid lines in (A) show the critical lines of the cluster at source redshifts of 1 and 7, respectively. Panels
(B), (C), and (D) zoom into three GGSL events enclosing sources at redshifts 1.425, 4.996, and 3.753,
respectively. The white lines in those panels show the critical lines of the lenses at the corresponding source
redshifts. In (B) and (D), the background lensed sources are bluer than the foreground lensing galaxies.
In (C), the lensed source is not visible in the HST image but is detected in an observation with the Multi-Unit
Spectroscopic Explorer (MUSE) spectrograph on the VLT  \cite{methods}. The source is detected at a wavelength of
$\sim 7289$  angstroms corresponding to the redshifted  Lyman-$\alpha$ spectral line of hydrogen, at locations indicated by the cyan
contours. The white crosses indicate the positions of four multiple images of the source. Equivalent images
for all the other clusters are shown in figs. S1 to S3.}
%    \caption{{\bf Color-composite image of the central region of the galaxy cluster MACSJ1206}. The image combines HST observations in the bands F105W, F110W, F125W, F140W, F160W (red channel), F606W, F625W, F775W, F814W, F850LP (green channel), F435W and  F475W (blue channel). The dashed and solid lines in panel A show the lens critical lines at source redshifts of 1 and 7, respectively. Panels B, C, and D zoom over three GGSL events enclosing sources at redshifts 1.425, 4.996, and 3.753. The white lines show the lens critical lines for the corresponding source redshifts. In panels A and B the lensed sources are bluer than the lenses. In panel C, the lensed source is not visible in the HST image. We discovered this system in an observation with the Multi-Unit-Spectroscopic-Explorer (MUSE) spectrograph of the VLT \cite{methods}. The source is visible at the wavelength of $\sim 7289$ angstroms, corresponding to the redshifted Lyman-$\alpha$ hydrogen spectral line, as shown by the cyan contours. The white crosses indicate the positions of four multiple images of the source. Equivalent images for all the other clusters we analyse are shown in Figs. S1-S3.}
\label{fig:macs1206_insets}
\end{figure}

\begin{figure}
    \centering
    \includegraphics[width=\textwidth]{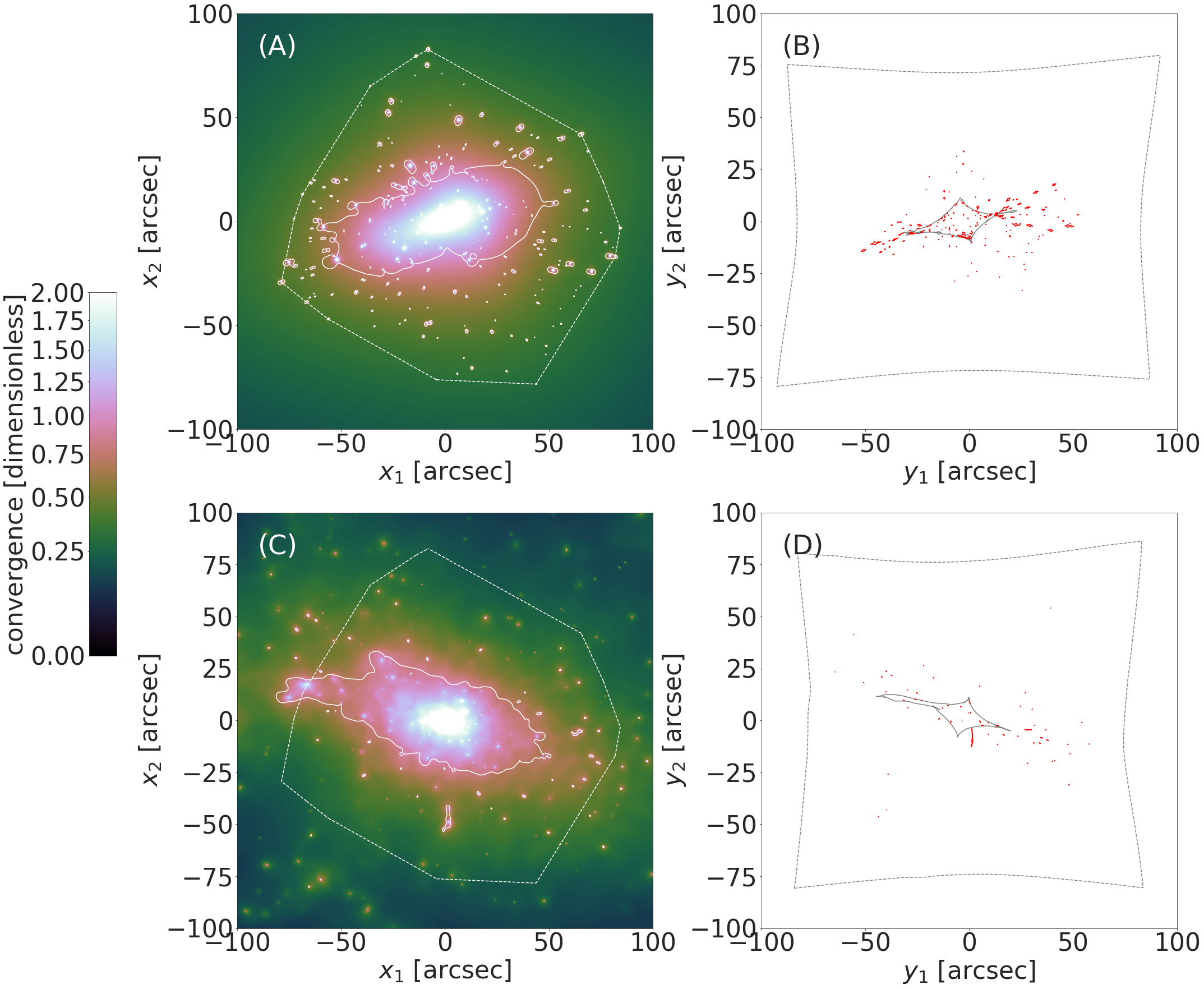}
    \caption{{\bf Comparison between an observed and a simulated gravitational lens: (A)} The projected mass map (called convergence) of MACSJ1206 (color bar), overlaid with the critical lines for sources at redshift $z=7$ (solid white lines). The dashed polygon delimits the region of the HST image within which cluster galaxies were selected and included in the lens model. {\bf (B)} The caustics corresponding to the principal (in gray) and to the secondary critical lines (in red) of MACSJ1206 \cite{methods}. The dashed gray line shows the limits of the field-of-view in (A) mapped into the source plane (12). The GGSL probability is calculated by dividing the area of the secondary caustics by that enclosed by the dashed gray line. {\bf (C)} The projected mass map and the critical lines for sources at redshift $z=7$ of a simulated cluster with a mass similar to that of MACSJ1206 \cite{methods}. The dashed polygon is the same as in (A). {\bf (D)} Caustics of the simulated cluster shown in (C). Although the main critical lines and caustics have similar extents, the secondary critical lines and caustics are larger and more numerous in the lens model of MACSJ1206 than in the simulation.}
%    \caption{{\bf Comparison between an observed and a simulated gravitational lens:} Panel A shows the projected mass map of MACSJ1206 overlaid with the critical lines for sources at redshift $z=7$ (solid white lines). The dashed polygon delimits the region of the HST image within which cluster galaxies were selected and included in the lens model.  Panel C shows the mass map of a simulated cluster with a mass similar to MACSJ1206 \cite{methods}. Although the main critical lines have similar extensions, the number and sizes of the secondary critical lines are larger in the lens model of MACSJ1206. Panels B, D show the caustics. Those corresponding to the principal critical lines are shown in grey, while the secondary caustics are shown in red \cite{methods}. The dashed grey line shows the limits of the field-of-view in panels A, B mapped into the source plane \cite{methods}. The GGSL probability is calculated by dividing the area of the secondary caustics by that enclosed by the dashed grey line.}
    \label{fig:simobs_examples}
\end{figure}

\begin{figure}
    \centering
    \includegraphics[width=\textwidth]{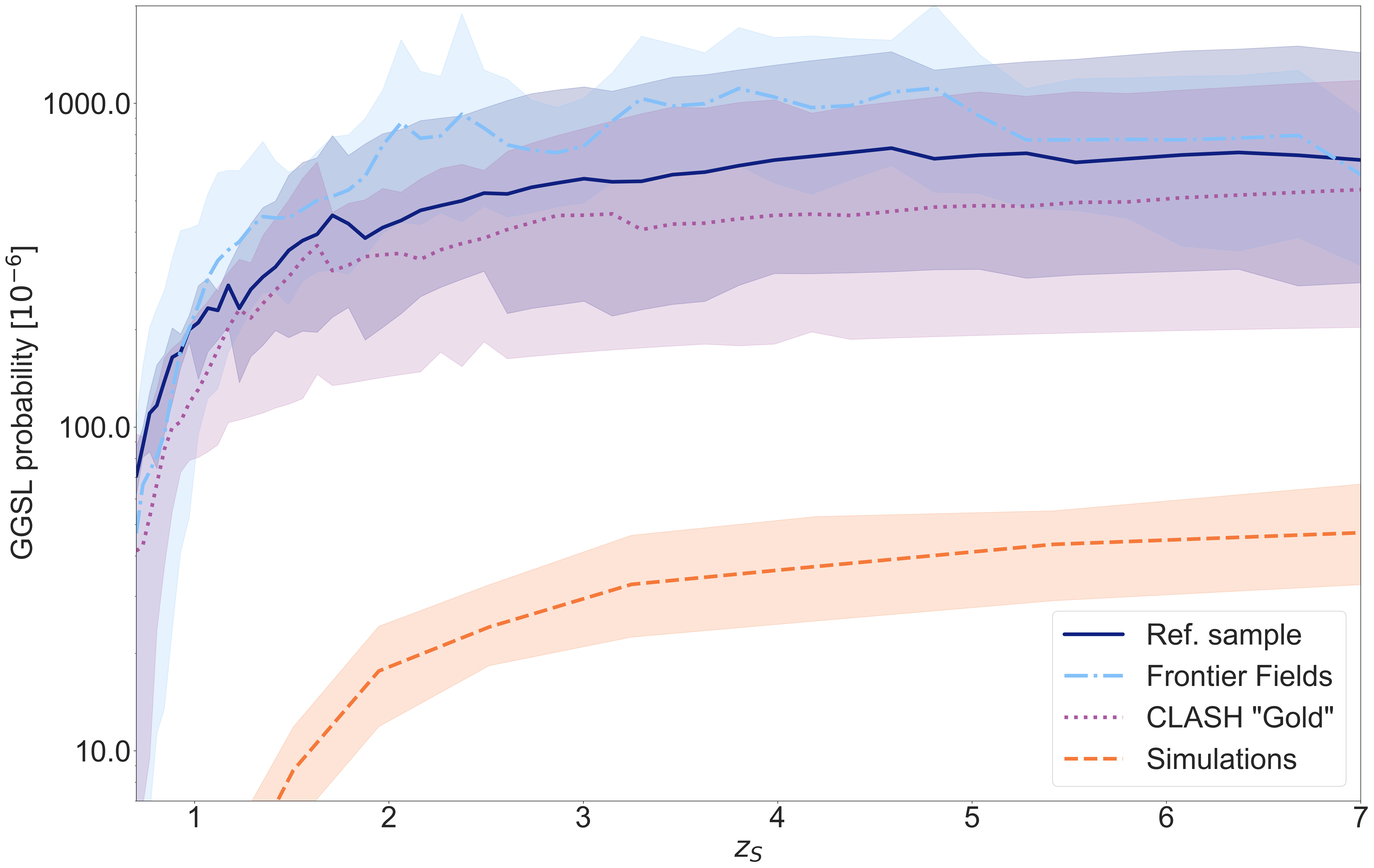}
    \caption{{\bf The GGSL probability as a function of the source redshift:} The mean GGSL probability among our reference sample is shown with a solid blue line. The light blue dot-dashed and violet dotted lines plot the computed GGSL probability for the HFF and CLASH Gold samples. The median GGSL probability measured from simulations is given by the orange dashed line \cite{methods}. The colored bands show the 99.9\% confidence intervals for each dataset. The discrepancy between observations and simulations is about an order of magnitude. }
    \label{fig:macs1206}
\end{figure}

\begin{figure}
    \centering
    \includegraphics[width=1\textwidth]{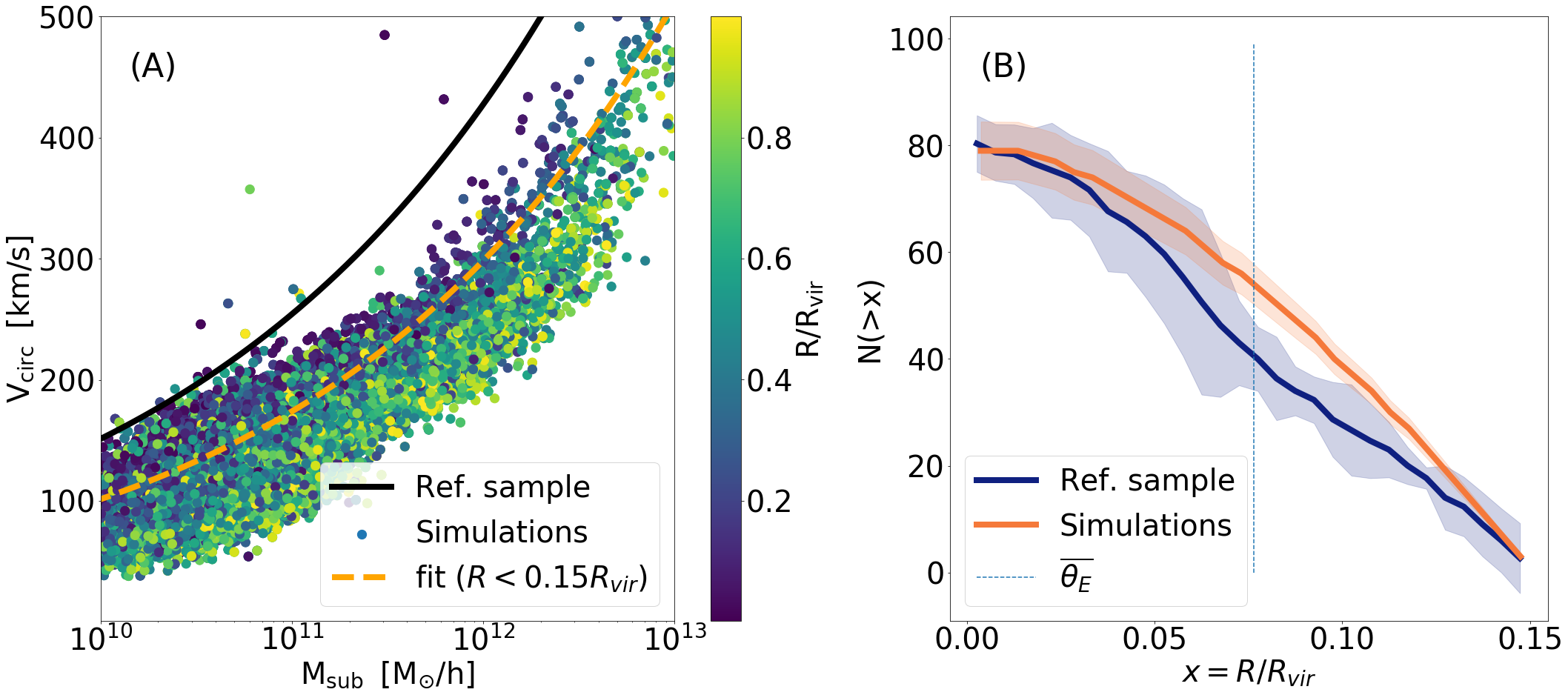}
    \caption{{\bf Circular velocities and positions of substructures in simulated and observed galaxy clusters: (A)} Substructure circular velocity as a function of substructure mass $M_{sub}$. The circular velocity is a proxy for the concentration of the substructure mass. The solid black line shows the average relation for the reference sample \cite{2019A&A...631A.130B}. The colored circles show the simulations, color-coded by the substructure distance from the cluster center $R$ in units of the cluster virial radius $R_{vir}$. The orange dashed curve shows the best-fit model relation for simulated substructures whose distance is less than 15\% of the virial radius. This is roughly the region around the cluster center probed by strong lensing. The observed relation is always above that derived from the simulations, indicating that observed substructures are more compact than the simulated ones. {\bf (B)}: mean cumulative distribution of the substructure distances from the cluster center, $N(>x)$. The distances are scaled by the virial radius of the host cluster, $x=R/R_{vir}$. The red and grey lines show the distributions for the observed reference sample and the simulations, respectively. The vertical dashed line shows the mean size of the main critical lines of MACSJ1206, MACSJ0416, and AS1063, $\overline\theta_E$.}
    \label{fig:vm_rd}
\end{figure}

\section*{Acknowledgments}
We thank S. White and F. van den Bosch for insightful discussions. We also thank G. Murante for sharing with us the numerical simulations used in this work and Al. Benitez-Llambay for making public his code \textsc{pysphviewer}.

\section*{Funding}
This work was performed in part at Aspen Center for Physics, which is supported by National Science Foundation grant PHY-1607611. We acknowledge support from the Italian  Ministry of Foreign Affairs and International Cooperation, Directorate General for Country Promotion, from PRIN-MIUR 2015W7KAWC, from PRIN-MIUR 2017WSCC32, from PRIN MIUR 2017-20173ML3WW\_001, from ASI through the Grant ASI-INAF n. 2018-23-HH.0 and ASI-INAF n.2017-14-H.0, from INAF (funding of main-stream projects), and from the INFN INDARK grant. P.N. acknowledges support from the Aspen Center for Physics for the workshop titled "Progress after Impasse: New Frontiers in Dark Matter" that she co-organized in Summer 2019 and the Space Telescope Science Institute grant HST-GO-15117.021. C.G. acknowledges support by VILLUM FONDEN Young Investigator Programme through grant no.~10123. SB acknowledges financial support from the EU H2020 Research and Innovation Programme under the ExaNeSt project (Grant Agreement No. 671553).

\section*{Author contributions}
MM coordinated the project, performed the lensing analysis of the simulated clusters, measured the lensing cross sections and probabilities of both simulated halos and observed clusters, contributed to the modeling of the observed clusters. GD developed the algorithm to measure the lensing cross sections. PB, PR, GBC, AM, and CG built the strong lensing models and the spectroscopic catalogs of MACSJ1206, AS1063, MACSJ0416, and of the “Gold” CLASH sample. CG performed the \textsc{MOKA} simulations of MACSJ1206 and analyzed the sub-halo catalogs of the simulated clusters. PN, FC, and EV contributed to the analysis of the simulations and to the interpretation of the results. ER and SB produced the numerical simulations and the sub-halo catalogs. RBM produced the multi-lens plane simulations used to test effects of matter along the line of sight. MM, PN, and FC wrote the manuscript including contributions from all the other authors.

\section*{Competing interests}
We declare no competing interests.

\section*{Data and materials availability} The lens models of all clusters in the reference and in the CLASH Gold samples are available as \textsc{Lenstool} parameter files at this URL: https://bit.ly/2AHnc2E. 
The mass models of the clusters in HFF sample can be downloaded as lens maps from this website: \\https://archive.stsci.edu/prepds/frontier/lensmodels/. We used the version v4 of the models produced by the teams CATS and Sharon.
The snapshot files of all simulated halos can be downloaded from this URL: https://bit.ly/2DeqXgC. The simulated subhalo catalogs are available at this URL: https://bit.ly/2AQZ1Pa.
The GLAMER software for ray-tracing and the code used to measure the GGSL cross sections are available at Zenodo \cite{glamer,GGSLcalculator}.

\newpage

\newpage
\section*{\underline{Supplementary materials}}

\setcounter{equation}{0}
\renewcommand{\theequation}{S\arabic{equation}}

\subsection*{Materials and methods}

\subsubsection*{Conventions}

The lens models for the observed galaxy clusters were built assuming a flat $\Lambda$-Cold Dark Matter cosmology with a matter density parameter at the present epoch of $\Omega_0=0.3$, and a Hubble parameter of $H_0=70$ km s$^{-1}$ Mpc$^{-1}$. For the simulated sample of clusters, slightly different values for the cosmological parameters were used, with $\Omega_0=0.24$  and a Hubble constant of $H_0=72$ km s$^{-1}$ Mpc$^{-1}$. As noted in the main text, we use abbreviations when referring to the various observed cluster samples (these abbreviations are reproduced in the caption for Table~\ref{table:clusters}).

\subsubsection*{Observational data-set}

The observational data-set used in this analysis consists of 11 galaxy clusters, listed in Table~\ref{table:clusters}. These are divided into three distinct samples. The first one, is our reference sample. It comprises clusters for which models of the mass distribution are based on the combination of deep HST (Hubble Space Telescope) imaging and spectroscopic measurements. In addition, for the reference sample measured stellar kinematic priors were also included to constrain the masses of the cluster members. This combination of imaging and spectroscopic data provides mass models with the highest fidelity for the reference sample. These models are published in \cite{2019A&A...631A.130B}.

The reference sample contains MACSJ1206 ($z=0.439)$,  MACSJ0416 ($z=0.397$), and AS1063 ($z=0.348)$. These three clusters were part of the CLASH program \cite{2012ApJ...749...97Z}, and were observed  with the Advanced Camera for Surveys (ACS, \cite{1998SPIE.3356..234F}) and Wide-Field Camera 3 instrument (WFC3, \cite{2011ApJS..193...27W}) cameras on the HST in 16 broad band filters, covering the wavelength range from the UV (Ultra-Violet) to near-IR (near-Infra Red). HST imaging of MACS0416 and AS1063 was augmented with data from the Hubble Frontier Fields (HFF) program \cite{2017ApJ...837...97L} yielding exposures in seven additional filters (F435, F606W, F814W, F105W, F125W, F140W, F160W). The HFF program, with data from about 140 orbits per cluster, provides the deepest images currently available for cluster fields. These three clusters also have data from ground-based telescopes and were part of a spectroscopic follow-up campaign with the Cluster Lensing And supernova Survey with Hubble - Very Large Telescope (CLASH-VLT) Large program (P.I. P. Rosati), that used the high-multiplexing spectrograph VIsible Multi-Object Spectrograph (VIMOS) \cite{2003SPIE.4841.1670L}. Additional spectroscopic information in the core region of these clusters was obtained using the Multi-Unit Spectroscopic Explorer integral field spectrograph (MUSE) also mounted on the VLT \cite{2010SPIE.7735E..08B}. MUSE has a field of view of 1 arcmin$^2$, a spatial sampling of $0.200"$, a spectral resolution of $\sim 2.4$\AA$ $ over the spectral range $4750-9350 $\AA$ $, and a spectral sampling of $1.25 $\AA$ $ pix$^{-1}$. The MUSE data were obtained as part of several Guest Observer programs that are listed in \cite{2019A&A...631A.130B}. A subset of redshift measurements of cluster member galaxies and multiple images used by \cite{2019A&A...631A.130B} were previosuly published by \cite{2017arXiv170700690C,2017A&A...600A..90C,2015A&A...574A..11K}.
Color-composite HST images of MACSJ1206, MACSJ0416, and AS1063 are shown in Figs.~1 and S1.

Two other samples are also included in the analysis for comparison. The second sample consists of the four remaining galaxy clusters observed in the HFF program \cite{2017ApJ...837...97L}. These are A2744 ($z=0.308$), MACSJ0717 ($z=0.545$), MACSJ1149 ($z=0.542$), and A370 ($z=0.375$). Each cluster was observed by HST with the filters F435, F606W, F814W, F105W, F125W, F140W, and F160W, adding up to a total exposure of 140 orbits per cluster. HST images of A2744, MACSJ0717, MACSJ1149, and A370 are shown in Fig.~S2. Unlike the reference sample, we do not build lens models for these clusters. Instead, we utilize the publicly available reconstructions downloaded from the Frontier Fields website. These models were produced by several independent groups using in addition to the HST imaging data described above, and spectroscopic measurements from several coordinated observing programs \cite{2014MNRAS.443.1549J,2014ApJ...797...48J,2015MNRAS.446.4132J,2015MNRAS.447.3130D,2015MNRAS.451.3920D,2015MNRAS.452.1437J,2016ApJ...819..114K,2016A&A...588A..99L,2017A&A...600A..90C,2017ApJ...842..132B,2017MNRAS.469.3946L,2018MNRAS.473..663M,2018MNRAS.473.4279D,2018MNRAS.480.3140W,2018ApJ...868..129S,2018ApJ...864...98B}.

Our final sample consists of published lens models \cite{2019arXiv190305103C} for a sample of four CLASH clusters with HST imaging and spectroscopy from MUSE and CLASH-VLT data. These clusters have deep ($>1$ hour) MUSE observations and each cluster has at least 5 spectroscopically confirmed sets of multiple images. We refer to these as the CLASH ``Gold" sample, comprising the clusters RXJ2129 ($z=0.234$), MACSJ1931 ($z=0.352$), MACSJ0329 ($z=0.450$), and MACSJ2129 ($z=0.587$). 
In Fig.~S3, we show the colour composite images of the four clusters in this sample.

\subsubsection*{Summary of lens modeling techniques}

The publicly available lens modeling software package {\tt Lenstool} \cite{1996ApJ...471..643K,2007NJPh....9..447J}, that uses parametric techniques for mass reconstruction \cite{2017MNRAS.472.3177M} was used to generate the cluster mass models used in this work. In {\tt Lenstool} the overall mass distribution of the cluster is modeled using the superposition of multiple self-similar components on large and small scales \cite{NAT97.2}. Each mass component is described by a set of parameters that characterize its density profile, shape, and orientation. Galaxy clusters are assumed to be composed of one or more smooth, large-scale halo components and multiple clumpy, small galaxy-scale, components as seen in cosmological simulations of structure formation and evolution. The spatial locations of the larger scale halo components and small scale sub-halos are chosen by assuming that light traces mass. The small scale dark matter sub-halos are assumed to be associated with the most luminous cluster galaxies as tracers of the overall cluster mass distribution. In these models, the presence of massive structures in the outer regions of the cluster are accounted for by the inclusion of an external shear term $\gamma_{\rm ext}$ and orientation $\theta_{\rm ext}$. Furthermore, to relate mass to light for the small galaxy-scale components of the cluster, a set of empirically motivated scaling relations are assumed to connect the characteristics of the mass profile with observable quantities like the galaxy luminosity. In this scheme, all model parameters are varied with the goal of accurately reproducing the locations of the observed families of multiple images. This is done by minimizing the separation between observed images and their model predicted locations.

\subsubsection*{Lens models: Reference Sample}

The previously published lens models for the clusters in our reference sample are described elsewhere  \cite{2019A&A...631A.130B}. We outline some key aspects here. These mass models are built once again, by combining large-scale and galaxy-scale components, both of which are modeled with a dual-Pseudo-Isothermal-Elliptical (dPIE) mass distribution  \cite{2005MNRAS.356..309L,2007arXiv0710.5636E}. The projected surface mass density $\Sigma$ of a dPIE is given by:
\begin{equation}
\Sigma(R)=\frac{\sigma_0^2}{2G}\frac{r_{\rm cut}}{r_{\rm cut}-r_{\rm core}}\left(\frac{1}{\sqrt{R^2+r_{\rm core}^2}}-\frac{1}{\sqrt{R^2+r_{\rm cut}^2}}\right) \;.
\end{equation}
In the equation above $G$ is Newton's constant and the parameters characterizing each mass component are the 1D-velocity dispersion $\sigma_0$; the core radius $r_{\rm core}$; and the cut radius $r_{\rm cut}$. We assume negligible core radii justified by observations of early type galaxies, whose lensing properties are consistent with isothermal inner slopes \cite{2007ApJ...667..176G,2009ApJ...703L..51K,2006ApJ...649..599K,2009MNRAS.399...21B,2011MNRAS.415.2215B,2010ApJ...724..511A,2018MNRAS.475.2403L}. To reduce the number of free parameters in the construction of the lens model, for the smaller scale components of the cluster, $\sigma_0$ and $r_{\rm cut}$ are related to the luminosity of the cluster galaxies via empirically derived scaling relations \cite{1997MNRAS.287..833N}:
\begin{eqnarray}
    \sigma_0 & = & \sigma_{0,\star}\left(\frac{L}{L_\star}\right)^{a} \;, \\
    r_{\rm cut} & = & r_{\rm cut,\star}\left(\frac{L}{L_\star}\right)^{b} \;.
    \label{eq:scaling}
\end{eqnarray}
Therefore, the entire population of cluster galaxies can be described by four parameters - the calibrating values $\sigma_{0,\star}$ and $r_{\rm cut,\star}$, and slopes $a$, $b$, which can further be reduced to three free parameters by assuming additional scaling of the mass-to-light ratio $M/L$ with luminosity, thereby imposing a relation between the slopes $a$ and $b$ \cite{2007ApJ...668..643L,2015MNRAS.447.1224M}. 
 
To model cluster galaxies, we adopt these scaling relations and include dynamical information for these cluster members derived from measurements of the central stellar velocity dispersions with the MUSE spectrograph \cite{2019A&A...631A.130B}. The measured velocity dispersions of several cluster member galaxies are used as prior information for the scaling relations above \cite{2019A&A...631A.130B}. The lens models of MACSJ1206, MACSJ0416, and AS1063 were built using $82, 102$ and $55$ multiple images respectively. The mass model for the core of MACSJ1206 is built with three large-scale and 265 smaller galaxy-scale components. The parameters are constrained with 82 spectroscopically confirmed multiple images arising from 27 strongly lensed background sources, spanning a wide range of source redshifts and distances from the cluster center \cite{2017arXiv170700690C}. The velocity dispersions of $\sim 60$ cluster galaxies were used as prior information for the scaling relations. In the case of MACSJ0416, the model includes two primary large scale dark-matter halos, centered on the two brightest cluster galaxies (BCGs), and includes 193 galaxy-scale components. Additional information derived from stellar kinematics measurements of 49 cluster galaxies was used as a prior to fit the scaling relations of the cluster members \cite{2019A&A...631A.130B}. The model of the cluster AS1063 is also built using two large scale smooth dark matter halos, one centered on the BCG and the second is used to mimic an asymmetric mass distribution in the North-West region of the cluster. The clumpy component of the cluster consists of 221 smaller scale mass distributions associated with cluster members and these are included in the model with priors from stellar kinematic measurements of 37 galaxies from MUSE data. 

\subsubsection*{Lens models: The other cluster samples}

The mass distributions for all the Frontier Fields clusters have been reconstructed using {\tt Lenstool} by two independent groups \cite{2014MNRAS.443.1549J,2014MNRAS.444..268R,2014ApJ...797...48J,2016A&A...588A..99L,2015MNRAS.452.1437J,2017MNRAS.469.3946L,2018MNRAS.473..663M,2019MNRAS.485.3738L}. The models show small differences, likely arising from the incorporation of slightly different input constraints and variations in modeling choices (e.g. usage of different families of multiple images, different selection of cluster members, different assumptions on the exponents and normalization of the scaling relations). We use version 4 of these publicly available lens models produced by the two teams. The detailed models are available on the HFF website {\footnote{https://archive.stsci.edu/prepds/frontier/lensmodels/}}.

The mass models of the clusters in the CLASH ``Gold" sample are based on fewer observed lensing constraints compared to the other two samples. They consist of $\sim 20-40$ multiply imaged sources per cluster, however with extensive spectroscopic information. This is due to the shallower survey strategy and lower masses of the CLASH lenses. Including these less well constrained mass models in our analysis allows us to extend and explore our analysis to a slightly lower lensing efficiency regime. This more modest regime is less prone to selection biases \cite{2014ApJ...797...34M}. CLASH clusters (including MACSJ1206) have been shown to be less affected by many of the typical biases that cluster lenses suffer from, such as those arising from orientation, concentration, and dynamics \cite{2010A&A...519A..90M}. In addition, the CLASH “Gold” sub-sample  and our reference sample differ in two other key ways, for instance, no spectroscopic prior on the velocity dispersion of the cluster members is used to build the models for the CLASH "Gold" clusters. While for our reference sample it is assumed that cluster galaxies lie on the fundamental plane, i.e.  their mass-to-light ratio is $M/L \propto L^{-\delta}$ with $\delta=0.2$, for the CLASH sample \cite{2019arXiv190305103C} it is assumed that the $M/L$ ratio is independent of the galaxy luminosity.

\subsubsection*{Numerical simulations}

We compare a mass-matched sample of simulated clusters to the reference sample of observed clusters. We refer to the simulated sample as the AGN set. The properties of this simulated AGN set are described in detail in other papers \cite{2014MNRAS.438..195P,2014ApJ...797...34M,2017MNRAS.467.3827P}. Here, we provide a brief summary of their characteristics.  Cluster halos for the AGN set were identified in a low--resolution periodic simulation box with co-moving size of 1 $h^{-1}$ Gpc in a flat $\Lambda$CDM model universe with present matter density parameter $\Omega_{m,0}=0.24$ and baryon density parameter $\Omega_{b,0}=0.04$. The Hubble constant adopted was $H_0=72$ km s$^{-1}$ Mpc$^{-1}$ and the normalisation of the matter power spectrum on a scale of $8\,h^{-1}$ Mpc was  $\sigma_8=0.8$. The adopted primordial power spectrum of the density fluctuations was $P(k) \propto k^{n}$ with $n=0.96$. The parent simulation followed 1024$^{3}$ collision-less (dark matter only) particles in the box.  The clusters were identified at $z=0$ using a standard  Friends-of-Friends (FoF) algorithm, and their Lagrangian region was re-simulated at higher resolution employing the Zoomed Initial Conditions code \cite{1997MNRAS.286..865T,2011MNRAS.418.2234B}. The resolution is progressively degraded outside this region to save computational time while still accurately describing the large--scale tidal field. The Lagrangian region was taken to be large enough to ensure that only high-resolution particles were present within five virial-radii of the clusters.

The re-simulations were then carried out using the TreePM--SPH {\small GADGET--3} code \cite{SP05.1,2016MNRAS.455.2110B} that now includes baryons. Several versions of the cluster halos are re-simulated in multiple runs that include variants of sub-grid models for baryonic physics. In one version, the runs include collisionless dark matter particles and baryons, the effects of radiative gas cooling, star formation, metal enrichment of the intra-cluster medium (ICM) and energy feedback from supernova explosions and Active Galactic Nuclei (AGN). To assess the effects of feedback on our final results, we also generate a version of the simulations where all other baryonic physics as noted above are implemented but feedback from the AGN is turned off (see Sect.\ref{sect:effofbaryons}). Star formation and stellar feedback were implemented according to the model proposed in \cite{2003MNRAS.339..289S}. Heating/cooling from cosmic microwave background (CMB) and from a UV/X-ray time-dependent uniform ionizing background are modeled as in\cite{2001cghr.confE..64H}. Rates of radiative cooling follow the prescription by \cite{2009MNRAS.393...99W}. The model for including Super-Massive-Black-Hole (SMBH) accretion and AGN feedback is described in \cite{2015MNRAS.448.1504S}. 

The dark matter and gas particle masses in these simulations are $8.47\times 10^8~h^{-1} M_\odot$ and $1.53\times 10^8~h^{-1} M_\odot$, respectively. For the gravitational force, a Plummer-equivalent softening length of $\epsilon = 3.75 h^{-1}$ kpc is used for DM and gas particles, whereas $\epsilon = 2 h^{-1}$ kpc is used for black hole and star particles. We consider six simulation snapshots in the redshift range $0.25 < z_{L} < 0.55$. At each redshift, the sample contains 25 simulated clusters that are the most massive halos identified in the original dark matter only parent cosmological box at redshift zero, that are mass-matched to the observed samples of cluster-lenses. The masses of the clusters are $M_{200} > 5\times 10^{14} h^{-1}M_\odot$. To augment the simulated sample size, we project each halo along three independent lines-of-sight, corresponding to the orthogonal axes of the simulation box.

\subsubsection*{Computation of the deflection angle maps}

To compare the lensing properties of the simulated clusters to observations, we compute deflection angle maps from their projected mass distributions via ray-tracing through the lens, starting from the position of the observer. First, we project the particles belonging to the halo along the desired line of sight on the lens plane to compute a mass map. To select particles, we define a slice of the simulated volume around the cluster, corresponding to a depth of $20 h^{-1}$Mpc. We need to achieve sufficient spatial resolution to resolve the smallest substructures in the simulated clusters. However, we also have to simultaneously minimize the noise due to the discreteness of the density field of the lens, which is described by a finite number of particles \cite{2013MNRAS.430.2232R}. This requires smoothing the particle mass distribution on scales of a few times the local mean inter-particle separation. The various particle species: dark matter, gas, stars, black holes, are spatially distributed in different ways. For example, most star particles are confined to the central regions of sub-halos, while dark matter particles are more  broadly distributed. To ensure adequate resolution for small scale structures, we employ a Smoothed-Particle-Hydrodynamics scheme to estimate the three-dimensional density around each particle. This smoothing is adaptive, in the sense that the smoothing length of each particle corresponds to the distance to its $n_{\rm smooth}$-th closest neighbouring particle in 3D space. This smoothing is done separately for each particle species. In other words, we produce smoothed density fields for each particle type and then finally sum them, projecting onto the lens plane. The described smoothing procedure is implemented using the python package {\tt py-sphviewer} \cite{alejandro_benitez_llambay_2015_21703}. We use $n_{\rm smooth}=50$ and obtain the composite mass distribution which can then be used to compute deflection angle maps. This is done using Fast-Fourier-Transform techniques (e.g. \cite{BA94.1,2010A&A...514A..93M}). Starting from the position of the virtual observer, we trace light-rays through a regular grid of $4096 \times 4096$ points covering a region of $400" \times 400"$ around the halo center on the lens plane. The pixel scale of this grid is thus $\sim 0.1"$. At redshifts between $z=0.2$ and $z=0.5$, this angular scale of  $0.1"$ corresponds to physical scales of $0.3$ kpc and $0.6$ kpc, respectively. This scale is sufficiently small to resolve galaxy-scale halos well spatially. This scale over-samples the scale of gravitational softening of the simulations.

\subsection*{Metric for comparing observational data with simulations}

\subsubsection*{Galaxy-Galaxy Strong Lensing cross section}
\label{sect:ggsl}
With the availability of these extremely well constrained high-fidelity mass models for observed cluster lenses, we are now in the position to devise metrics that will enable testing their small-scale properties with that of simulated clusters. As a metric to compare the observations to the simulations, we devise the Galaxy Galaxy Strong Lensing (GGSL) cross section, a property that provides an estimate of strong lensing events produced on small scales. In these GGSL events, an individual cluster member embedded in the larger scale cluster environment, acts as a lens for a background galaxy source. When multiple images of this lensed source are produced by the cluster galaxy lens, the separation of these multiple images is roughly the size of the Einstein radius of the individual galaxy acting as lens (i.e. scales of a few arcsec). To produce image splitting, the cluster galaxy lens must be super-critical (i.e. its surface mass density needs to exceed a critical value), so it generates a distinct critical line. For any lens, this condition for criticality is given by the equation:
\begin{equation}
\kappa(\vec x) \pm \gamma(\vec x)=1 \;,
\label{eq:critlines}
\end{equation}
where $\vec x$ indicates the position on the lens plane, and $\kappa$ and $\gamma$ are the convergence and the modulus of the shear \cite{SC92.1}. The plus sign ($+$) in equation (4) corresponds to the condition for the formation of the tangential critical line, in contrast to the radial critical line, which corresponds to the minus sign ($-$).  

{For each cluster shown in Figs.~S1, S2, and S3, we plot the computed critical lines for sources at redshifts $z_s=1$ and $z_s=7$ superimposed on the corresponding HST image  (a larger view of the critical lines of MACSJ1206 is also shown in Fig.~1).}  A galaxy cluster generally produces large scale critical lines arising from the smooth cluster scale component. These are not implicated in GGSL events, however, their spatial extent determines the ability of smaller scale sub-halos around individual cluster galaxies to generate their own critical lines. It is more common to satisfy the condition for criticality and to form small-scale critical lines in the immediate proximity of the cluster-scale critical lines, where the combination of $\kappa \pm \gamma$ is close to unity already. Thus, the larger the overall cluster's critical lines, the higher the probability that sub-halos in their proximity are also super-critical. We expect that galaxies in clusters can more easily be super-critical than in the field environment, as these lie in the higher surface-density background provided by the larger scale cluster component compared to field galaxies. The large scale smooth mass distribution boosts the lensing effect of the embedded smaller scale components.

Critical lines can be mapped into the corresponding caustics using the lens equation,
\begin{equation}
\vec y = \vec x -\vec{\alpha}(\vec x) \;,
\end{equation}
where $\vec{\alpha}(\vec x)$ is the lens deflection angle at the position $\vec x$. The deflection angle is obtained from the lens model of observed clusters and from the particle distributions of the simulated clusters as outlined above.

Observations of strong gravitational lensing by early-type galaxies show that the overall density profile of the lenses is nearly isothermal \cite{2009ApJ...703L..51K,2010ARAA..48...87T}. Under these circumstances, we expect the radial critical lines of massive cluster members (typically red galaxies) to be absent or very small (e.g. \cite{2011A&ARv..19...47K}, their figure 7). Therefore, we focus our attention on the tangential critical lines and  associated strong lensing features produced by cluster member galaxies.

We define the cluster GGSL cross section $\sigma_{\rm GGSL}$ as the area enclosed by all the galaxy-scale tangential caustics produced by cluster members, $\sigma_{GGSL}=\sum \sigma_{i}$, where the $\sigma_{\rm i}$ are the individual contributions. To compute the cross section, secondary critical lines need to be found and mapped onto the corresponding caustics on the source plane. We only consider the contribution to the cross section from those substructures that develop critical lines that are distinct from the primary, large scale, critical lines of the clusters. For example, the primary and the secondary critical lines of the galaxy cluster MACSJ1206 for sources at $z_s=7$ are shown in Fig.~S4. The cluster primary critical line has small scale wiggles and appendages. These features result from the merger of secondary and primary critical lines and do not contribute to the GGSL cross section, as defined.

We place a lower limit on the size of the galaxy-scale critical lines that we include in computing the GGSL cross section, so we can compute this exact quantity from our observational data. This lower limit is set to be 0.5 arcsec in equivalent Einstein radius (i.e. the radius of the circle with the same area as that of the critical line, \cite{2013SSRv..177...31M}). Note that this is $\sim$ 5 times larger than the resolution of HST images.

The GGSL cross sections are computed by summing the areas of all the secondary caustics. To compute these areas, we treat the caustics as sets of polygons. This summing procedure is implemented using the python package {\tt shapely} \cite{shapely07}. The critical lines are identified as level 0 contours in the map of $\lambda_t=1-\kappa-\gamma$ and mapped onto the corresponding caustics using the deflection angles.

\subsubsection*{Galaxy-Galaxy Strong Lensing probability}

Given that both the convergence and the shear change as a function the source redshift, the shape and extension of critical lines also vary with the redshift of the source. Consequently, the strong lensing cross section, and the GGSL cross section, depend on the source redshift. The scaling of the convergence and shear with source redshift is given by 
\begin{equation}
f(z_S|z_L) = \frac{D_{LS}(z_L,z_S)}{D_S(z_S)} \;,
\end{equation}
where $D_{LS}$ and $D_S$ are the angular diameter distances between the lens and the source planes and between the observer and the source plane, respectively. The growth of $\kappa$ and $\gamma$ as a function of the source redshift is fast for $z_S$ close to $z_L$ and then it slows down for increasingly higher values of $z_S$. The size of the critical lines grows in a similar manner (e.g. \cite{2017MNRAS.472.3177M}, figure 11). When a secondary critical line merges into the larger main critical line, we observe a decrement in the GGSL cross section.

The GGSL cross section divided by the size of the area on the source plane within the field-of-view (FOV) defines the probability of GGSL events,
\begin{equation}
    P_{\rm GGSL}(z_S)=\frac{\sum_i\sigma_i(z_S)}{A_s(z_S)} \;,
\end{equation}
where $\sigma_i$ is the area enclosed by the $i$-th secondary caustic, and $A_s$ is the FOV mapped on the source plane at $z_S$. The GGSL probability as a function of the source redshift derived for the observed clusters and the simulated sample are shown in Fig.~3. To compare with observations, we stack all the simulated halos and calculate the median GGSL probability as a function of the source redshift, as shown in Fig.~ 2. The result is plotted with the dashed orange line. The $99.9\%$ confidence limit around the median is quantified by bootstrapping the sample one thousand times.

\subsubsection*{Sub-halo masses and circular velocities}

From the simulations, we extract the distribution and characteristic properties of the substructure i.e. the sub-halos. The {\tt Subfind} code \cite{2001MNRAS.328..726S,2009MNRAS.399..497D} was used to decompose each cluster halo in our simulated data-set into sub-halos (e.g. \cite{2015ApJ...800...38G}). For each sub-halo, {\tt Subfind} provides a measurement of the maximum circular velocity (evaluated at correspondent radius $R_{max}$), $V_{\rm circ}=\sqrt{GM(R_{\rm max})/R_{\rm max}}$, of the total bound mass, $M_{\rm sub}$, and of the position. 

These derived sub-halo properties are compared to those of the sub-halos from the strong lensing models of the clusters in the reference sample. As noted previously, galaxy-scale sub-halos are included in the lens models as dPIE mass distributions with negligible core radii. Their mass profiles are fully characterised by the 1D-velocity dispersion $\sigma_0$ and by the cut-radius, $r_{\rm cut}$. The parameters follow the scaling relations in Eq.~\ref{eq:scaling}. The average relation between the total mass of the sub-halo $M_{\rm tot}$ and 1D-velocity dispersion is \cite{2019A&A...631A.130B}:
\begin{equation}
    M_{\rm tot}=3.5_{-0.9}^{+1.1} \times 10^{11}\,\rm M_{\odot}\,\left[\frac{\sigma_0}{220\,\rm(km\,s^{-1})}\right]^{4.43^{+0.02}_{-0.05}} \;.
    \label{eq:mass-sigma}
\end{equation}

Under the approximation that the galaxy profiles are singular and isothermal, the 1D-velocity dispersion can be converted into a circular velocity via $V_{\rm circ}=\sqrt{2}\sigma_0$.

\subsubsection*{Sub-halo mass functions}

Fig.~S5 shows the cumulative distribution of sub-halo masses, for sub-halos whose projected positions on the plane of the sky are within 15\% of the projected virial radius of the host cluster. This is approximately equivalent to twice the size of the average Einstein radius for the three clusters in the reference sample. We limit the analysis to sub-halos with total mass $M_{sub}>10^{10}\; h^{-1}\; M_\odot$. 

The cumulative mass functions derived from the strong lensing models are consistent with those of the simulations at the level of  $<2\sigma$. The sub-halo masses used in this analysis were measured using dynamical priors \cite{2019arXiv190305103C}, thus partially breaking the degeneracy with the other model parameters (i.e. the cut-radius). Similar results were obtained by \cite{2017MNRAS.468.1962N} without dynamical priors. 

\subsubsection*{The relation between sub-halo mass and velocity dispersion}

Fig.~4A shows the relation between the maximum circular velocity and mass for sub-halos derived from the lens models and the simulations. Fig.~4A also shows the relation in Eq.~\ref{eq:mass-sigma}, assuming $V_{circ}=\sqrt{2}\sigma_0$, and all the simulated sub-halos in the AGN dataset within the virial radius of their host. The closer a sub-halo is to the center, the larger its maximum circular velocity is at a given mass. In agreement with the previous findings \cite{2015ApJ...800...38G,2018ApJ...864...98B}, for a given mass, observed sub-halos have larger circular velocities compared to simulations, even if we consider only those at the closest distances from the cluster center. This indicates that their mass distributions are much more compact than their simulated analogs. This result holds over three-orders of magnitude in mass.

\subsubsection*{Spatial distribution of sub-halos}

Fig.~4B shows the cumulative projected distribution of sub-halos as a function of cluster-centric distance. We limit the analysis to sub-halos located within $0.15R_{vir}$ and with masses $M_{sub}>10^{10}\; h^{-1}\; M_\odot$. The cluster-centric distances are scaled by the virial radius of the cluster halo. As shown previously, we also find that sub-halos derived from observations tend to be slightly more abundant at small radii from the cluster center \cite{2017MNRAS.468.1962N}. Sub-halos aggregate toward the center of the cluster and also lie closer to the main critical lines of the overall cluster. Their enhanced intrinsic compactness and their proximity to the cluster critical lines likely boosts the sub-halo strong lensing cross sections. 

\subsection*{Validation tests}

We performed the following tests to validate both the simulations and the lens models and check the robustness of our results.

\paragraph{Number of expected GGSL events:}

{The GGSL cross section can be converted into an expected number of GGSL events. As the surface brightness is conserved in gravitational lensing, for a given number density of sources $n ({\rm S,z_s})$ with surface brightness $S$ above a given detection limit, the total number of GGSL events $N_{\rm GGSL}$ can be estimated as:
\begin{equation}
N_{\rm GGSL}=\int_{S_{lim}}^\infty \int_{z_L}^{\infty}n(S,z_S)\sigma_{GGSL}(z_S)d z_S d S \;,
\label{eq:nggsl}
\end{equation}
where $S_{lim}$ is the surface brightness limit. The source number density is estimated from deep surveys like the HUDF \cite{2006AJ....132.1729B}. The expected number of GGSL events in MACSJ1206 as a function of the limiting surface brightness is shown in  Fig.~S6. Using a photometric redshift catalog of the HUDF galaxies  \cite{2015AJ....150...31R} and the surface brightness measured in the F850LP filter, we find that the expected number of GGSL events up to $z_s=7$ in the field shown in Fig.~1 is $\sim 3$ at the depth of the HUDF. For the CLASH  images, which are shallower than those of the HFF, we expect this number to be smaller by a factor $\sim 2$. Given that the area of the HUDF is very small ($\sim 11$ sq. arcmin), our estimates are probably affected by cosmic variance, but there are no other data-sets that reach the depth of the HFFs that can be used to estimate the source number density out to highest redshifts. The relative cosmic variance for unclustered sources is $\sim40-60\%$ between redshift $\sim5$ and $\sim 3$\cite{2004ApJ...600L.171S}.  Based on similar calculations using the lens models of MACSJ0416 and AS1063, we expect a $\sim 1$ GGSL event in each of the two clusters. 

These are conservative, and are probably lower limits to the expected counts, because, at shallow magnitudes, the HUDF appears to be under-dense compared to wider fields such as those covered by the COSMOS survey \cite{2007ApJS..172..219L} \cite{2010PASP..122..947D}. In addition, our calculations do not account for magnified, unresolved sources.}

\paragraph{Finite source size effects:} 

{Some of the sources lensed in a GGSL event can have spatial extents that exceed the secondary caustics in size (e.g. Fig.~1C). Our procedure to compute the GGSL cross section and probability is valid only in the case of small source sizes.  Extended sources can also give rise to GGSL events even if they are only partially contained within a secondary caustic. This can be accounted for by using a buffer of the appropriate size around the caustic, when computing the contribution from the sub-halo to the cross section. For example, assuming circular sources of fixed radius $0.5"$, the GGSL cross sections and (and the expected number of events) would increase by a factor $\sim 3$ for both the simulations and the lens models. The discrepancy between the GGSL cross-sections, however, remains.

We test the mismatch between simulations and observations by plotting the number density of secondary caustics as a function of the source redshift instead of the computed GGSL probability. The results of this test are shown in Fig.~S7, which illustrates that the number of secondary caustics in simulated clusters is also inconsistent with the observations of the clusters in our samples. The lens models of these observed samples cluster samples produce more critical lines and caustics around cluster galaxies than found in numerically simulated halos.}

\paragraph{Effects of limited resolution:} 

The mass and spatial resolution of our simulations are sufficient to resolve the internal structure of the substructures. Re-simulating a sub-sample of the clusters increasing the mass resolution by a factor of 10 and correspondingly increasing the force resolution, we found that the GGSL probability remains largely unchanged as shown in Fig.~S8.  

We also produced mock realizations of the clusters in our sample using the semi-analytic code \textsc{MOKA} \cite{2012MNRAS.421.3343G}, that uses combinations of parametric mass distributions and inputs from numerical simulations to generate lens mass models. \textsc{MOKA} reproduces the mass and radial distribution functions of substructures in numerically simulated clusters. The density profiles of both the host halos and the sub-halos are modeled analytically, thus allowing us to reach arbitrary mass and spatial resolutions. Substructures are modeled using truncated singular-isothermal profiles. 

In Fig.~S9A, we show the comparison between the GGSL probability in MACSJ1206 and in lens models generated with 127 independent realizations of \textsc{MOKA} (Case 1). The generated MOKA models match the total mass and the concentration of MACSJ1206 \cite{2016ApJ...821..116U}. The results are consistent with those found by comparing observations of MACSJ1206 and the full numerical simulations of the AGN sample. There is a discrepancy between the \textsc{MOKA} models and the MACSJ1206 lens model now more than $\sim 1$ order of magnitude.  

\paragraph{Comparison with other simulations}

{We compared our simulations to others with different force and mass resolutions, that in addition, also implement alternative feedback models. The Illustris and the Hydrangea simulations \cite{2017MNRAS.470.4186B}, have mass resolutions that are 2-3 orders of magnitude better than the simulations described above. The Illustris simulations \cite{2014MNRAS.444.1518V} are limited to a smaller cosmological box (106.5 Mpc on a side) and, for this reason, they contain only lower mass halos (the highest mass halo in the box has a total mass of $\sim 1.3\times 10^{14}\; M_\odot$). The Hydrangea simulations (which are part of the ``Cluster-Eagle project'' \cite{2017MNRAS.471.1088B} and use the galaxy formation model developed for the EAGLE simulations (Evolution and Assembly of GaLaxies and their Environments) \cite{2015MNRAS.446..521S}) are a suite of zoom-in simulations of massive clusters, similar to our simulations. We derive the median relation between the maximum circular velocity and the mass of the sub-halos in the Hydrangea clusters from figures 10 and 11 of \cite{2017MNRAS.470.4186B}. In Fig.~S10, which is a modified version of Fig.~4A, the Hydrangea results and the measurements for the sub-halos in the 13 most massive groups/clusters  in the Illustris-1 simulation box (using snapshots 105 and 108, corresponding to redshifts $z_L=0.46$ and $z_L=0.4$) are shown. Although the simulations do not perfectly overlap, they are consistently below the $V_{max}-M_{sub}$ relation that we measure in our reference sample. This demonstrates that sub-halos extracted from numerical simulations performed by independent groups, using different set-ups and codes, all have maximum circular velocities for sub-halos that are smaller than found in observations.}

\paragraph{Artificial disruption of sub-halos:}

We investigated how numerical effects might lead to the artificial disruption of sub-halos and hence impact our conclusions. It has been argued that these effects could cause the under-estimation of the normalization of the sub-halo mass function in CDM simulations by up to a factor $\sim 2$ \cite{2019MNRAS.490.2091G}. We test how the GGSL probability as defined, could be affected by such a deficit of sub-halos using the code \textsc{MOKA}. As done earlier for verifying the results of the simulations as a function of spatial resolution, we produce constrained realizations of the mass distribution for the galaxy cluster MACSJ1206, correcting the sub-halo mass function by a factor 2 to take the under-estimate due to artificial disruption into account (Case 2 of the \textsc{MOKA} models). Fig.~S9A shows that, in this case, the GGSL probability also increases by a factor $\sim 2$, but this is still insufficient to bring the simulations into agreement with the model of MACSJ1206 given that the gap is about an order of magnitude. This reinforces our conclusion that the discrepancy that we have found is mainly due to the compactness of the sub-halos - a marked difference in their internal structure - and not their abundance.

As an additional test, we modified the spatial distribution of sub-halos in the \textsc{MOKA} simulations to be more concentrated towards the center of the cluster compared to the original model, as found in some simulations which include baryons \cite{2017MNRAS.466.4974M} (Case 3 of the \textsc{MOKA} models). Baryons make the sub-halos more resistant to tidal stripping and increase the effects of dynamical friction. Results show that even accounting for a more centrally concentrated distribution of sub-halos and the possible disruption of sub-halos due to numerical effects (Case 4 of our \textsc{MOKA} models), the mismatch between \textsc{MOKA} models and the reconstruction of MACSJ1206 remains at the level of an order of magnitude, as shown in Fig.~S9A.

\paragraph{How realistic is the simulated cluster sample?} We test whether the simulations used in our study are representative of the observed cluster sample and are hence not biased. The masses and redshifts of the simulated halos are in the range of the observed clusters.

The simulated and the observed samples also share the property of being in a variety of dynamical states. They are composed of a mixture of relaxed and unrelaxed systems. MACSJ1206 and the clusters in the CLASH “Gold” sample were part of the CLASH ``relaxed” cluster sample, while some the Frontier Fields clusters are merging systems. A description of the dynamical state of the simulated clusters and various statistical measures of equilibrium  can be found in \cite{2015ApJ...813L..17R} and \cite{2016ApJ...827..112B}. \cite{2018ApJ...864...98B}. Multi-wavelength studies of the three clusters in our reference sample and their different mass components \cite{2017ApJ...842..132B,2018ApJ...864...98B}. The three clusters in the reference sample appear to be in different dynamical states, with hot gas over total mass in the inner regions that differ amongst them, but in conjunction they reflect the diversity of the simulated clusters well. 

As noted, for our comparison between observed and simulated clusters we match them by virial mass, however, there is good agreement between the simulations and the observations also in terms of central surface mass density profiles. The surface density profiles of the simulated halos and observed clusters used in this work are shown in panel A of Fig.~S11. In fact, our simulations show a large diversity of central surface mass density profiles, which encompass those measured in the data.  The observed profiles tend to populate the upper part of the figure, indicating that there might be some residual orientation bias, i.e. the observed clusters, in particular those in the reference and FF samples, might be oriented with their major axes pointing towards us, thus producing higher central surface densities. This is an expected selection bias as observed dramatic cluster lenses tend to be those preferentially oriented toward us along the line of sight.
As shown in panel B of Fig.~S11 we find that the discrepancy between observations and simulations persists even when the contribution of each substructure to the total GGSL cross section is weighted by the inverse of the surface mass density of the larger scale component at the substructure position. This indicates that the discrepancy cannot be explained by the observed clusters having a larger surface mass density than the simulated ones. It also remains when simulated halos are matched to observed clusters using the size of their Einstein radii (e.g. Fig.~2). The size of the Einstein radius of the overall cluster measures its lensing strength, which is enhanced due to orientation, over-concentration, or dynamical biases \cite{TO04.2,2010A&A...519A..90M}.

We also compared the projected halo concentration distributions of the simulated clusters with the observational sample. The concentrations are obtained by fitting the surface mass density profiles with NFW models (Navarro-Frenk-White), defined as $c_{200}=r_{200}/r_s$, where $r_{200}$ is the radius enclosing a mean density equal to 200 times the critical density of the universe\cite{NA97.1}. The projected ellipticities of the simulated halos cover a broad range of values and are also consistent with observations. Comparison of the concentration distributions and the ellipticity distributions for the simulated and observed sample are shown in panels C and D of Fig.~S11. 

In addition, the agreement between our simulated sample and the observations in terms of stellar masses of the BCGs and of other massive galaxies has been discussed previously \cite{2018MNRAS.479.1125R,2019A&A...630A.144B}. The stellar mass function in the Hydrangea simulation suite also matches the observations \cite{2017MNRAS.470.4186B}. Other works [e.g. \cite{2014MNRAS.438..195P,2015ApJ...813L..17R,2018MNRAS.474.4089T}] also show that the simulations reproduce many of the key observed properties of clusters, such as X-ray scaling relations, radial profiles of entropy and density of the intra-cluster gas.

\paragraph{How important are the effects of baryons?}
\label{sect:effofbaryons}

Our results are robust even when feedback from SMBH accretion is turned off in the simulations. Energy and momentum feedback from black holes suppress star formation in substructures, altering the slope of their inner density profiles, making them less centrally concentrated and hence, weaker gravitational lenses. While the detailed physics of these AGN feedback processes is not well understood, without its inclusion, simulations are known to suffer the over-cooling problem \cite{2003MNRAS.344..835B}. Without feedback, a large amount of gas would be converted into stars at the center of halos and sub-halos, forming unrealistically dense cores, thus enhancing the ability of substructures to act as strong lenses. Such enhanced star formation is not seen in observed cluster galaxies. In  Fig.~S9B, we compare the GGSL probabilities of the reference AGN sample to those obtained from the same clusters simulated without AGN feedback. We refer to this sample as `no AGN' sample. The GGSL probability of these halos is indeed higher, and inches closer to that of observed clusters. However, these simulations would be grossly discrepant from observations in the total fraction of cluster baryons converted into stars.

\paragraph{How important are structures along the line-of-sight for GGSL?}

We also investigate the possibility that the GGSL probability could be enhanced by uncorrelated matter along the LOS to these clusters. Including the deflection of light by multiple lens planes generated using inputs from large cosmological simulations, we find that substructure critical lines and caustics on the smaller scales are not affected by matter along the LOS, as shown in Fig.~S12. {The GGSL cross section of real galaxy clusters only accounts for the contribution from galaxies whose cluster membership has been determined either spectroscopically or photometrically. In the case of simulated clusters we only consider the contribution from substructures in the same lens plane as the cluster. This lens plane is obtained by projecting the mass within a box centred on the cluster halo and a depth of $20 h^{-1}$ Mpc. Therefore, we caution that we might be inadvertently including substructures that are outside the cluster virial radius in the GGSL cross section calculation. The LOS structures are simulated by assuming that the rest of the light-cone, excluding the cluster, is an unbiased representation of the Universe.  It is populated with halos/galaxies according to a Sheth-Tormen mass function \cite{SH02.1}. This prescription does not take into account the clustering of halos so the variation between fields is likely larger, but we consider this adequate for our purposes.}

\newpage

\subsection*{Supplementary tables}

\setcounter{table}{0}
\renewcommand{\thetable}{S\arabic{table}}

\begin{center}
\begin{table}[ht!]
\begin{tabular}{|| c | c | c | c | c || }
 \hline\hline
 Cluster name & RA & Dec & Redshift & Sample \\ [0.5ex] 
 \hline\hline
 MACS J1206.2-0847  & 12 06 12.200 & -08 48 2.00 & 0.439 & Ref.\\ 
 Abell S1063  & 22 48 54.300 & -44 31 7.00 & 0.348 & Ref. \\  
 MACSJ 0416.1-2403  & 04 16 8.380 & -24 04 20.80 & 0.397 & Ref. \\
 \hline
 Abell 2744  & 00 14 20.030 &  -30 23 17.80 & 0.308 & FF \\
 MACS J0717.5+3745  & 07 17 36.500 & +37 45 23.00 & 0.545 & FF \\
 MACS J1149.5+2223  & 11 49 35.800 & +22 23 55.00 & 0.542 & FF \\
 Abell 370  & 02 39 50.500 & -01 35 8.00 & 0.375 & FF \\
 \hline
 RX J2129.7+0005  & 21 29 40.500 & +00 05 47.00 & 0.234 & CLASH ``Gold'' \\
 MACS J1931.8-2635  & 19 31 49.600 & -26 34 33.00 & 0.352 & CLASH ``Gold'' \\
 MACS J0329.7-0211  & 03 29 41.600 & -02 11 47.00 & 0.450 & CLASH ``Gold'' \\
 MACS J2129.4-0741  & 21 29 26.000 & -07 41 28.00& 0.587 & CLASH ``Gold'' \\
 \hline\hline
\end{tabular}
\caption{List of observed galaxy clusters used in this work. First column: cluster name; Second and third columns: Celestial coordinates [Right-Ascension (RA) and Declination (Dec) relative to the epoch January 1, 2000 (J2000)]; Fourth column: Redshift; Fifth column: abbreviated cluster sample name (Ref.:reference sample from \cite{2019A&A...631A.130B}; FF: Hubble Frontier Fields \cite{2017ApJ...837...97L}; CLASH ``Gold'': sub-sample of CLASH clusters from \cite{2019arXiv190305103C}.}
\label{table:clusters}
\end{table}
\end{center}

\newpage
\subsection*{Supplementary figures}

\setcounter{figure}{0}
\renewcommand{\thefigure}{S\arabic{figure}}

\begin{figure}
    \includegraphics[width=1.0\textwidth]{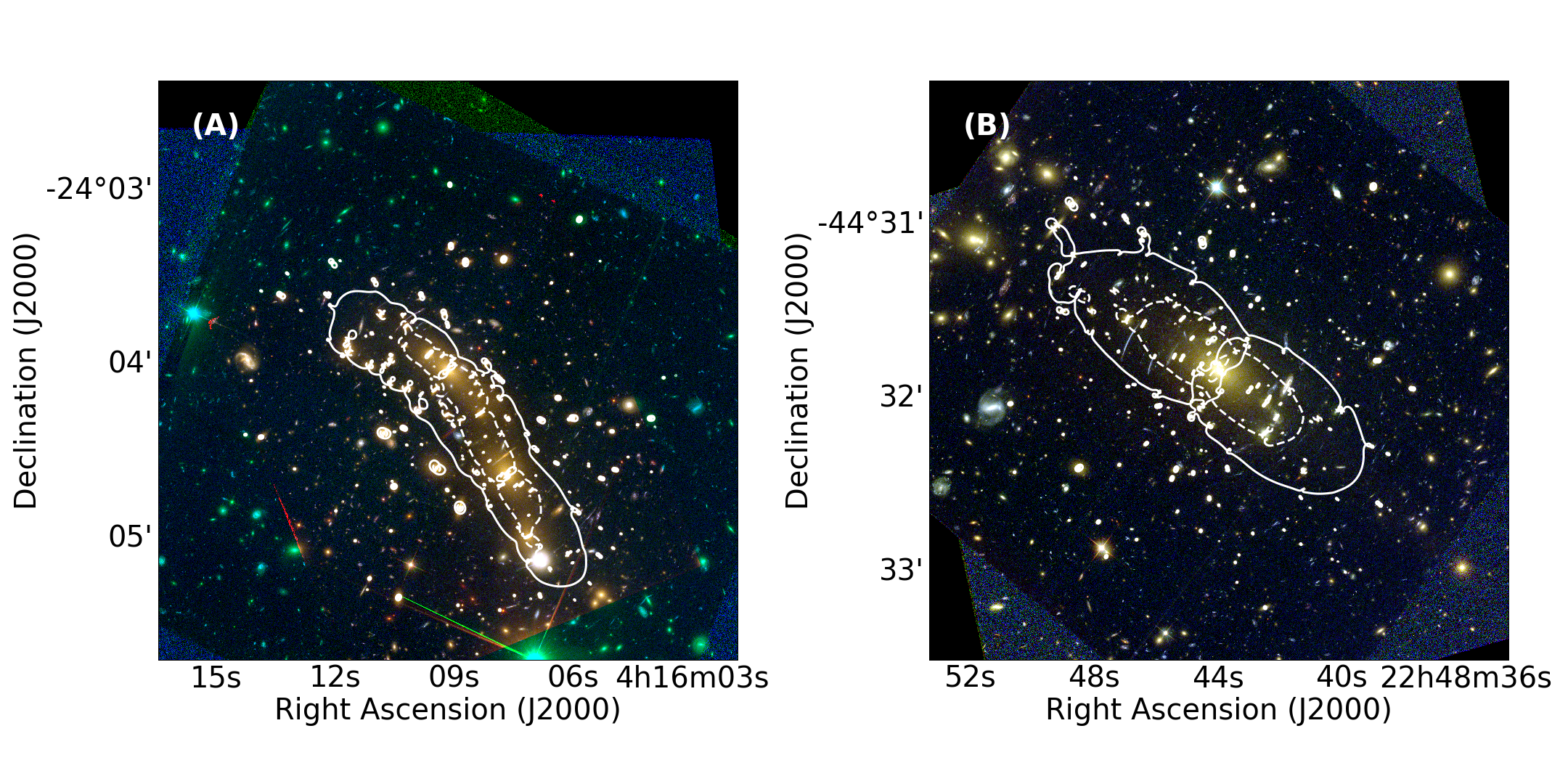}
    \caption{{\bf Galaxy clusters in the reference sample:} Same as Figure 1, but for clusters MACSJ0416 (panel A) and AS1063 (panel B).  Each panel corresponds to an area of $200\times200$ arcsec.}
    \label{fig:reference_sample}
\end{figure}

\begin{figure}
    \includegraphics[width=1.0\textwidth]{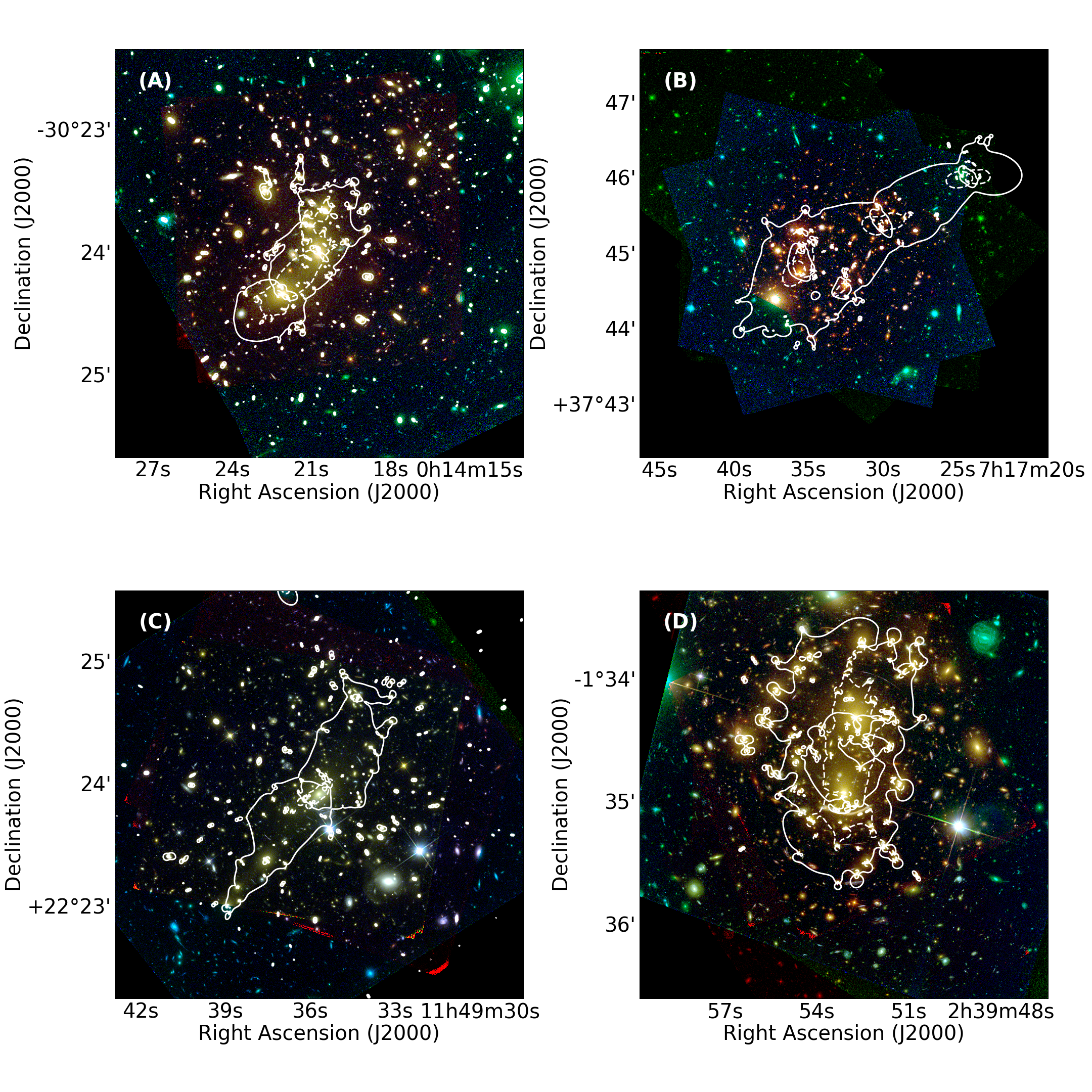}
    \caption{{\bf Galaxy clusters in the FF sample:} Same as Figure 1, but for the FF clusters A2744 (panel A), MACSJ0717 (panel B), MACSJ1149 (panel C), and A370 (panel D).  Each panel corresponds to an area of $200\times200$ arcsec, with the exception of the image of MACSJ0717 which covers an area that it four-times larger.}
    \label{fig:ff_sample}
\end{figure}

\begin{figure}
    \includegraphics[width=1.0\textwidth]{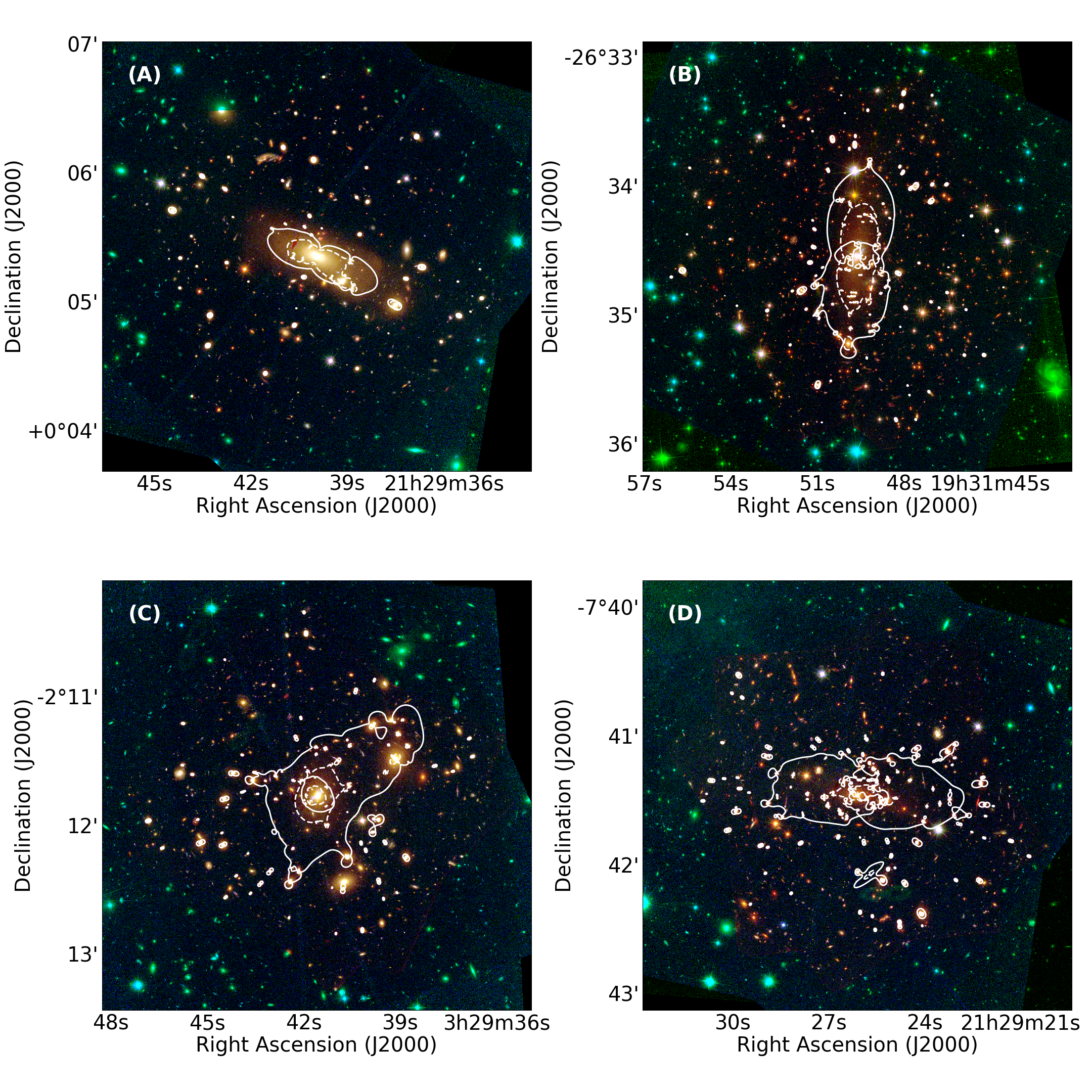}
    \caption{{\bf Galaxy clusters in the CLASH ``Gold" sample:}  Same as Figure 1, but for the clusters RXJ2129 (panel A), MACSJ1931 (panel B), MACSJ0329 (panel C), and MACSJ2129 (panel D).}
    \label{fig:clash_sample}
\end{figure}

\begin{figure}[tbp]
\centering 
\includegraphics[width=1.0\textwidth]{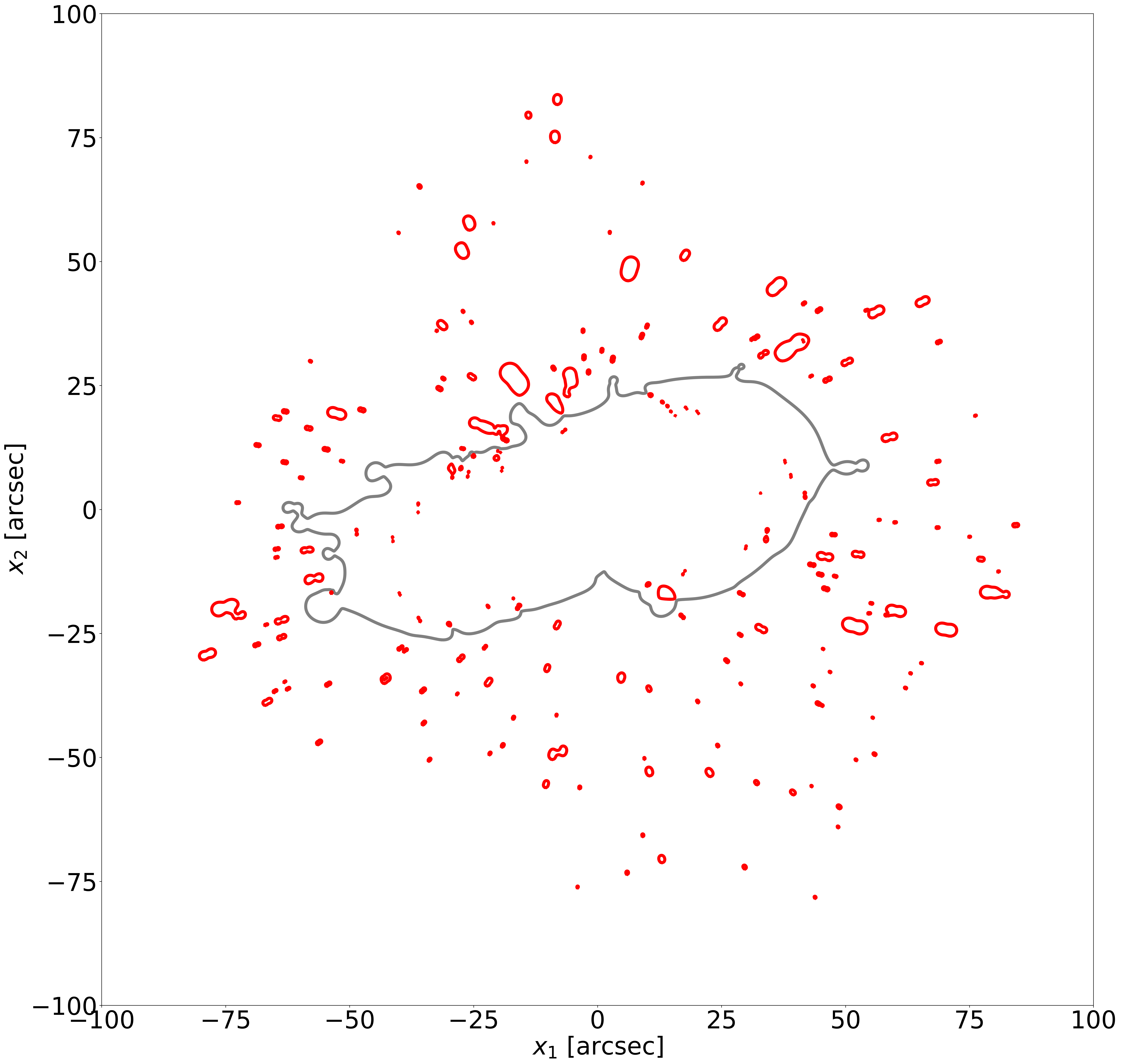}
\hfill
\caption{\label{fig:macs1206_crit_caust} {\bf Identification of primary and secondary tangential critical lines of the galaxy cluster MACSJ1206.} The primary critical lines are shown in gray, while the secondary critical lines are shown in red. The results in this figure are for a source redshift $z_S=7$. The corresponding caustics are shown in Figure 2B.}
\label{fig:crit_caust_m1206}
\end{figure}

\begin{figure}
\centering 
\includegraphics[width=\textwidth]{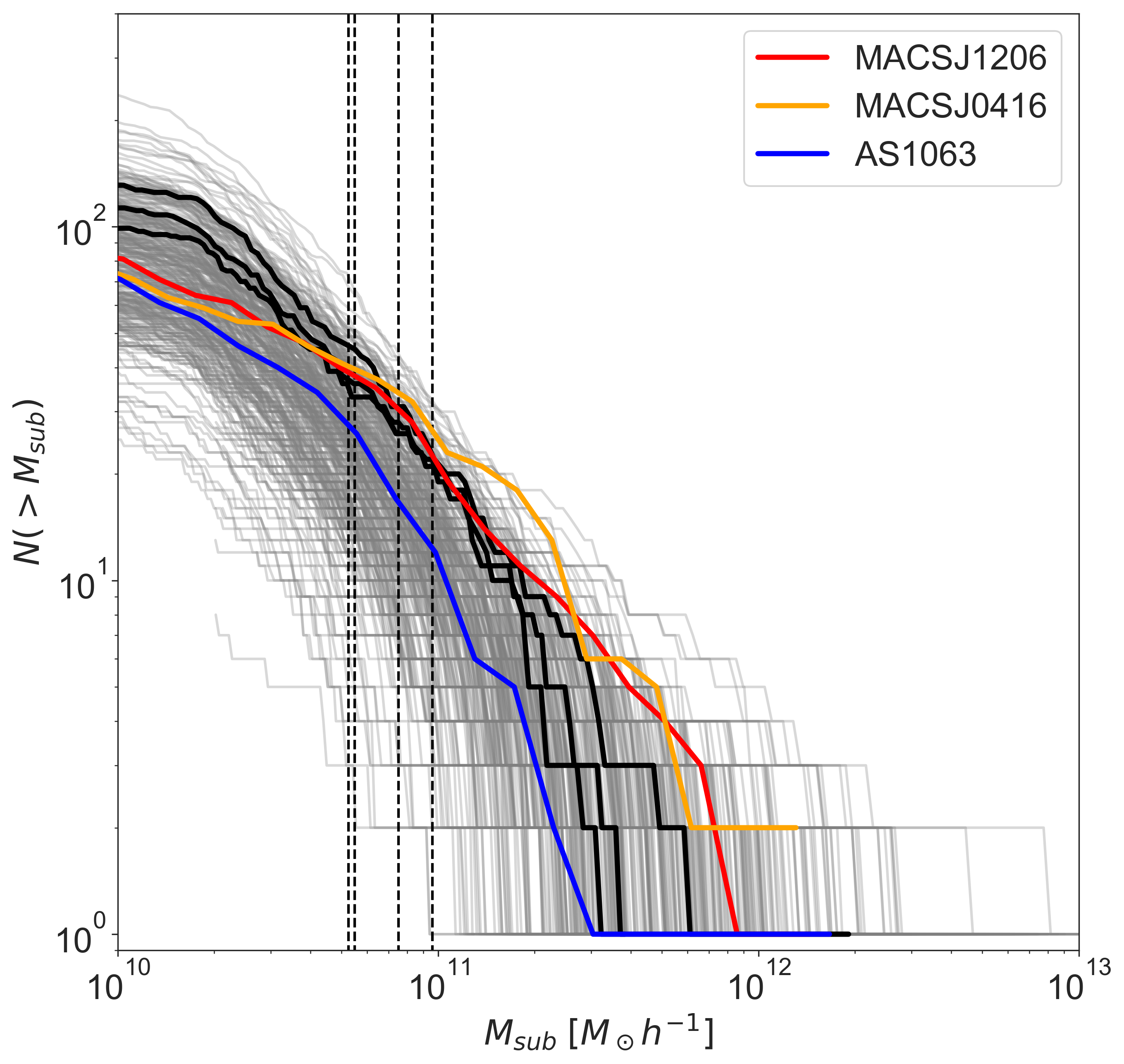}
\hfill
\caption{\label{fig:subhalos_pdf}
{\bf Cumulative distributions of the sub-halo masses. }
The grey lines show the results for three projections of each cluster in the AGN dataset, stacking all snapshots corresponding to $0.25 < z_{L} < 0.55$, within a cluster-centric distance of 0.15 $R_{vir}$ to match observations. The red, blue, and orange curves show the results from the lens models of MACSJ1206, MACSJ0416, and AS1063. {The vertical dashed lines indicate the masses of the three galaxy scale lenses found in MACSJ1206 (shown in Fig.~\ref{fig:macs1206_insets}) and of system ID14 in MACSJ0416 \cite{2017arXiv170700690C,2017ApJ...842...47V}. Three black solid lines are used to show the cumulative mass functions measured along three different lines-of-sight to one of the simulated clusters.} 
}
\label{fig:validation_tests}
\end{figure}

\begin{figure}
\centering
\includegraphics[width=1.0\textwidth]{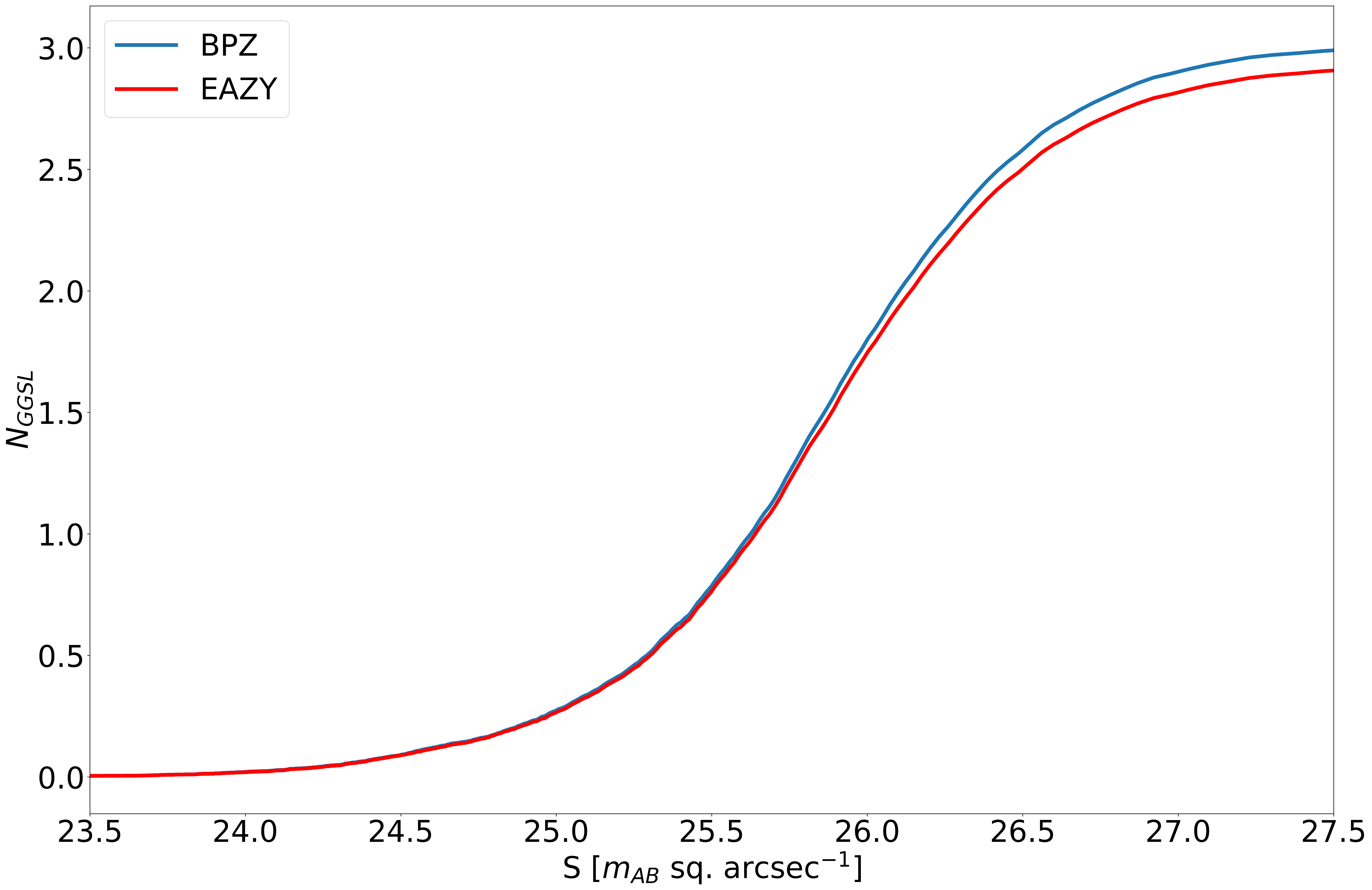}
\hfill
\caption{\label{fig:expected_ggsl} {\bf Expected number of GGSL events in the field of MACS1206.} The number is calculated using Eq.~\ref{eq:nggsl}, assuming the source number density $n(S,z)$ as measured in the HUDF, based on  photometric redshift catalogs  \cite{2015AJ....150...31R}. The blue and the red lines show the results based on the photometric redshift estimates obtained with two different codes, namely BPZ \cite{2011ascl.soft08011B} and EAZY \cite{2008ApJ...686.1503B}. The estimated number of events is shown as a function of the minimal source surface brightness.}
\end{figure}

\begin{figure}
\centering 
\includegraphics[width=1.0\columnwidth]{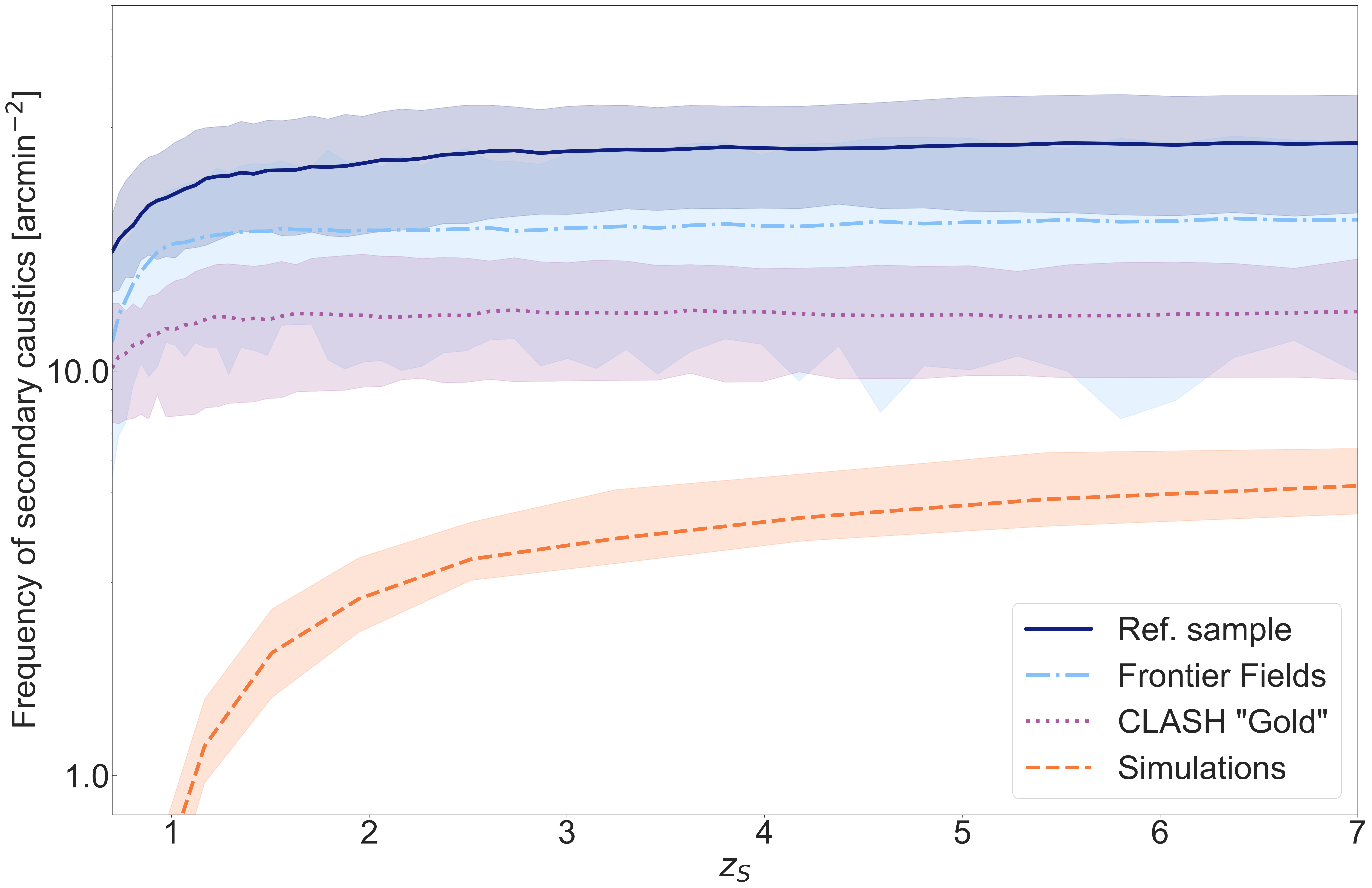}
\hfill
\caption{{\bf Comparison between spatial number densities of secondary caustics of simulated and observed galaxy clusters.} Colors and line-styles are the same as in Fig.~\ref{fig:macs1206}.}
\label{fig:comparesimobs_num}
\end{figure}

\begin{figure}
    \centering
    \includegraphics[width=1.0\columnwidth]{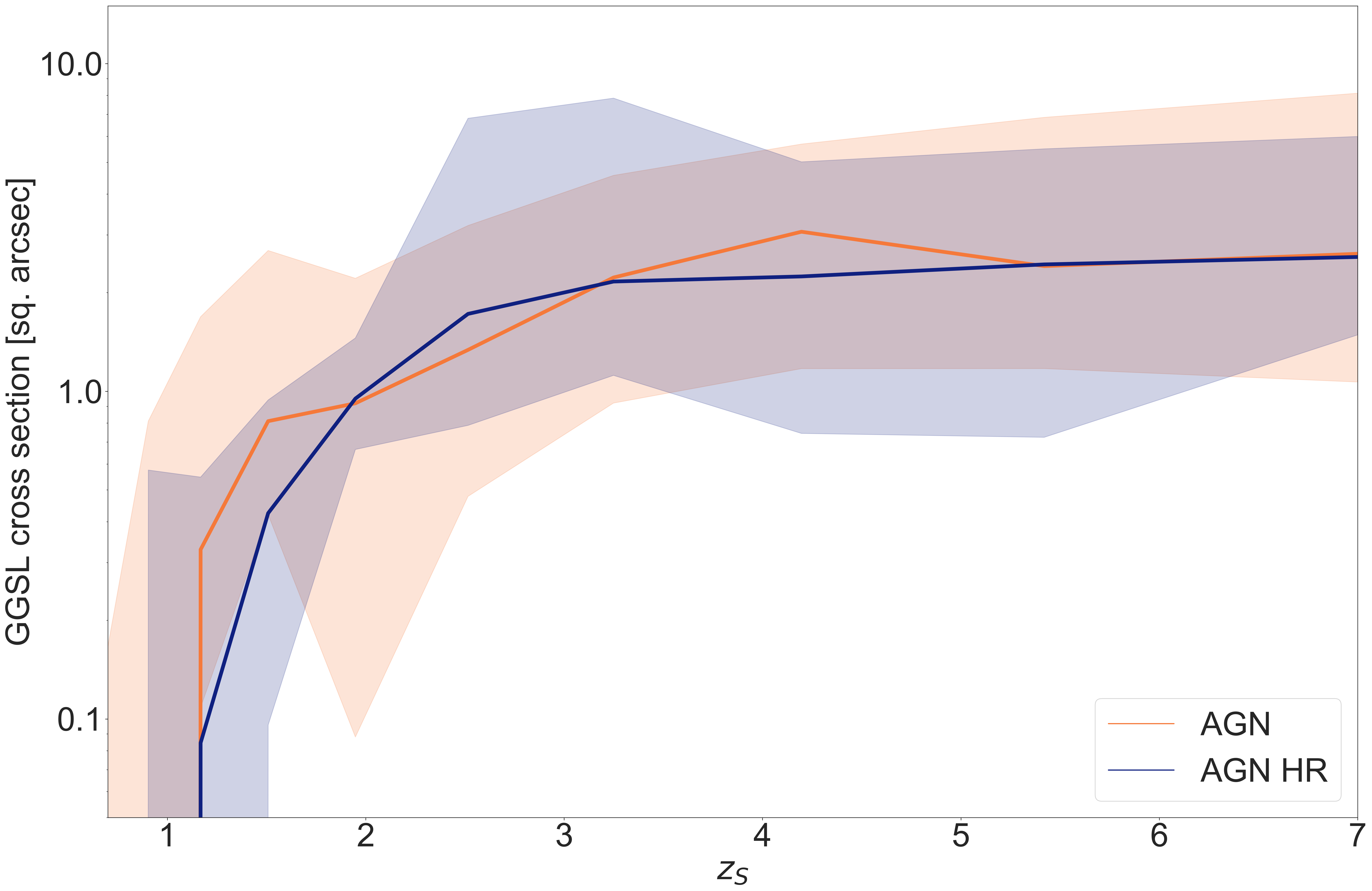}
    \caption{{\bf Effects of mass resolution:} The GGSL probability as a function of the background source redshift is shown for the same halo simulated at the same mass and force resolution as the AGN simulations (orange solid curve) and with a mass resolution ten times higher (AGN HR, blue solid curve).}
    \label{fig:HRvsSR}
\end{figure}

\begin{figure}
\centering 
\includegraphics[width=1.0\columnwidth]{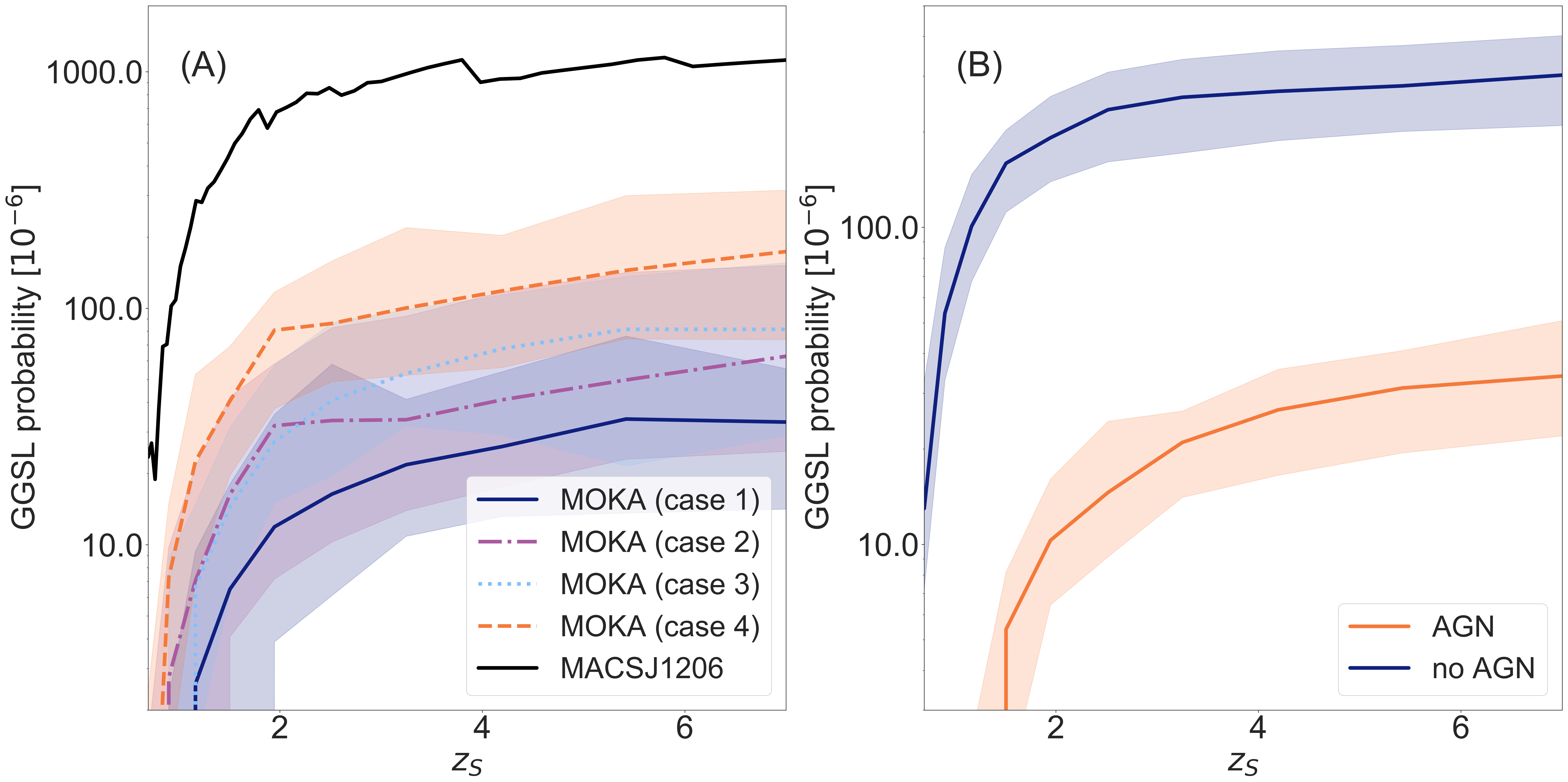}
\hfill
\caption{\label{fig:validation_tests1} {\bf Effects of resolution and AGN feedback:} Panel A: Probability for GGSL in MACSJ1206 (black curve) and in equivalent lens models generated with the code \textsc{MOKA} \cite{2012MNRAS.421.3343G}. The orange curve shows the results using the original recipes for the sub-halo mass  and radial distribution functions implemented in \textsc{MOKA} (Case 1). The dash-dotted violet curve shows the results obtained by increasing the normalization of the sub-halo mass function by a factor 2 to account for the numerical effects (Case 2). The dotted light blue curve shows the results obtained by modifying the radial distribution function of sub-halos, to better match simulations with baryons (Case 3). The orange dashed curve accounts for both a larger normalization of the mass function and a more centrally concentrated spatial distribution of sub-halos (Case 4). The colored bands show the $99.9$ confidence regions.
Panel B: Comparison between the probabilities for GGSL in the samples of clusters simulated with and without AGN feedback (orange and blue curves, respectively). The colored bands show the $99.9$ confidence regions.
}
\end{figure} 

\begin{figure}[ht!]
\centering 
\includegraphics[width=\textwidth]{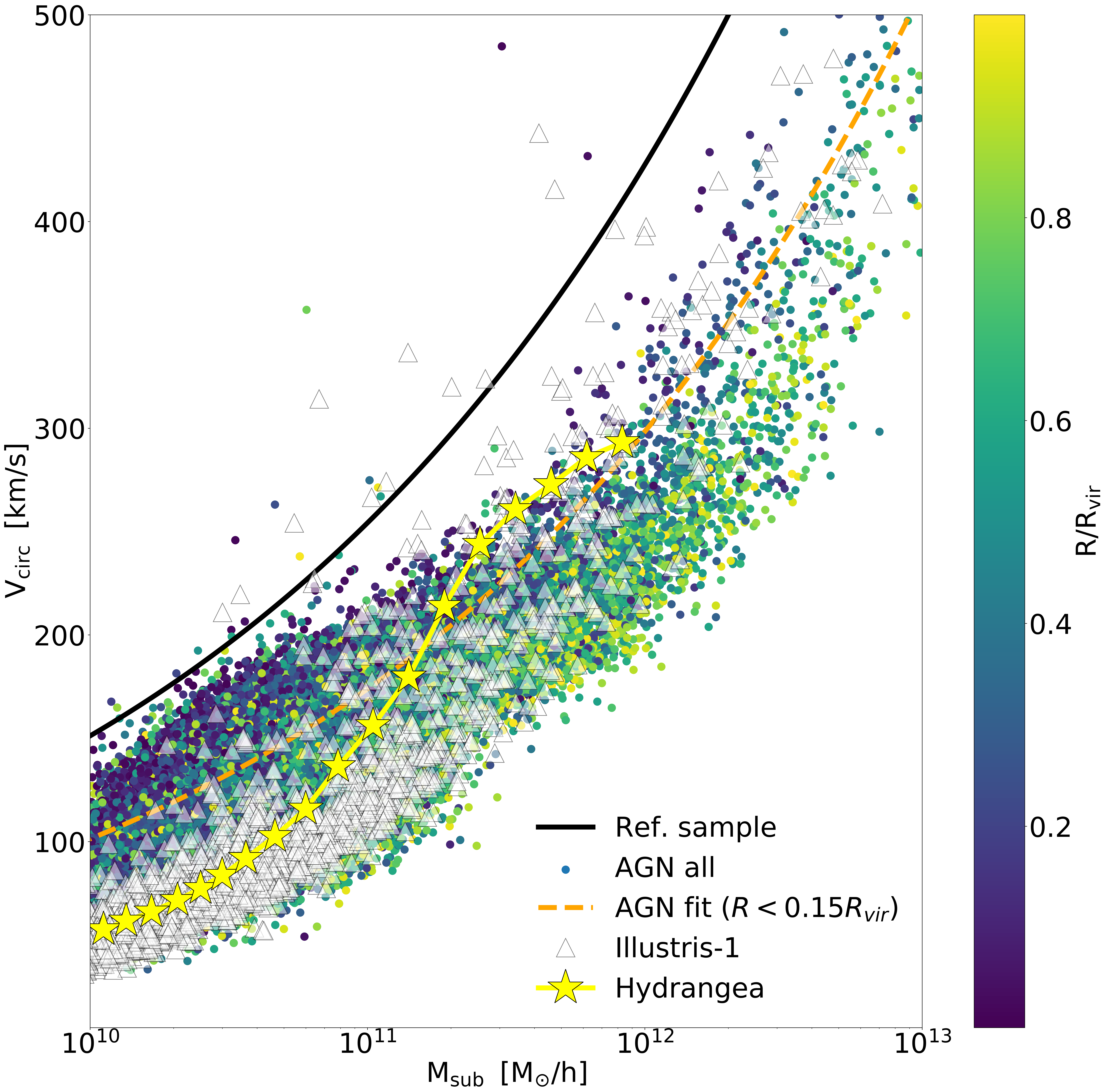}
\hfill
\caption{\label{fig:vmax_msub} 
{\bf Comparison with other numerical simulations.} Same as Fig 4A, but overlay now displays the relation between $V_{circ}$ and $M_{sub}$ from the Hydrangea simulations for massive cluster halos (yellow stars and solid line) \cite{2017MNRAS.470.4186B}. 
For comparison, we also show with white triangles the results for the sub-halos extracted from the 12 most massive halos in the Illustris-1 simulation at $z_L=0.4$ and $z_L=0.46$ \cite{2014MNRAS.444.1518V}. 
}
\end{figure}

\begin{figure}[ht!]
\centering 
\includegraphics[width=1\textwidth]{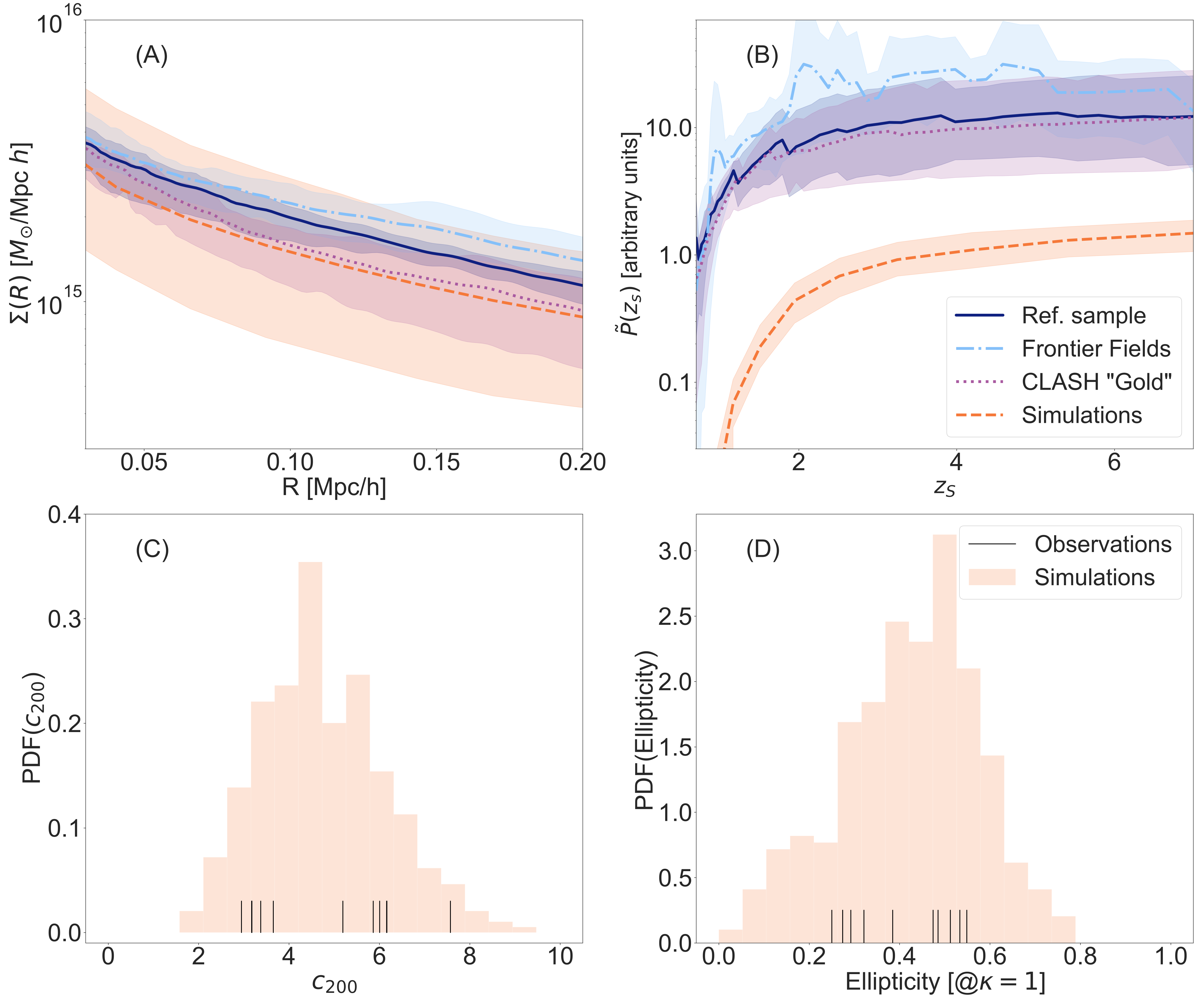}
\hfill
\caption{\label{fig:S10}
{\bf Comparison of observed and simulated cluster properties:} 
Panel A: Range of simulated and observationally determined central surface mass density profiles for the clusters. The orange dashed line is the median profile derived from the AGN simulations. The solid dark-blue, dash-dotted light blue, and dotted violet lines show the mean profiles of the clusters in the reference, FF, and CLASH ``Gold'' samples, respectively. The colored bands show the corresponding ranges between the first and the 99th percentiles of the profile distributions. Panel B: Same as Fig.~3 but for the GGSL probability computed by weighting the contribution of each substructure to the total GGSL cross section by the inverse of the surface mass density of the large scale component at the substructure position. Panel C: Distribution of the halo concentrations measured in the AGN simulations (orange histogram). The concentrations of the observed clusters in all samples are marked by the black sticks on the bottom. Panel D: same as panel C but for the ellipticities measured at the convergence level $\kappa=1$.}
\end{figure} 

\begin{figure}[ht!]
\centering 
\includegraphics[width=0.9\columnwidth]{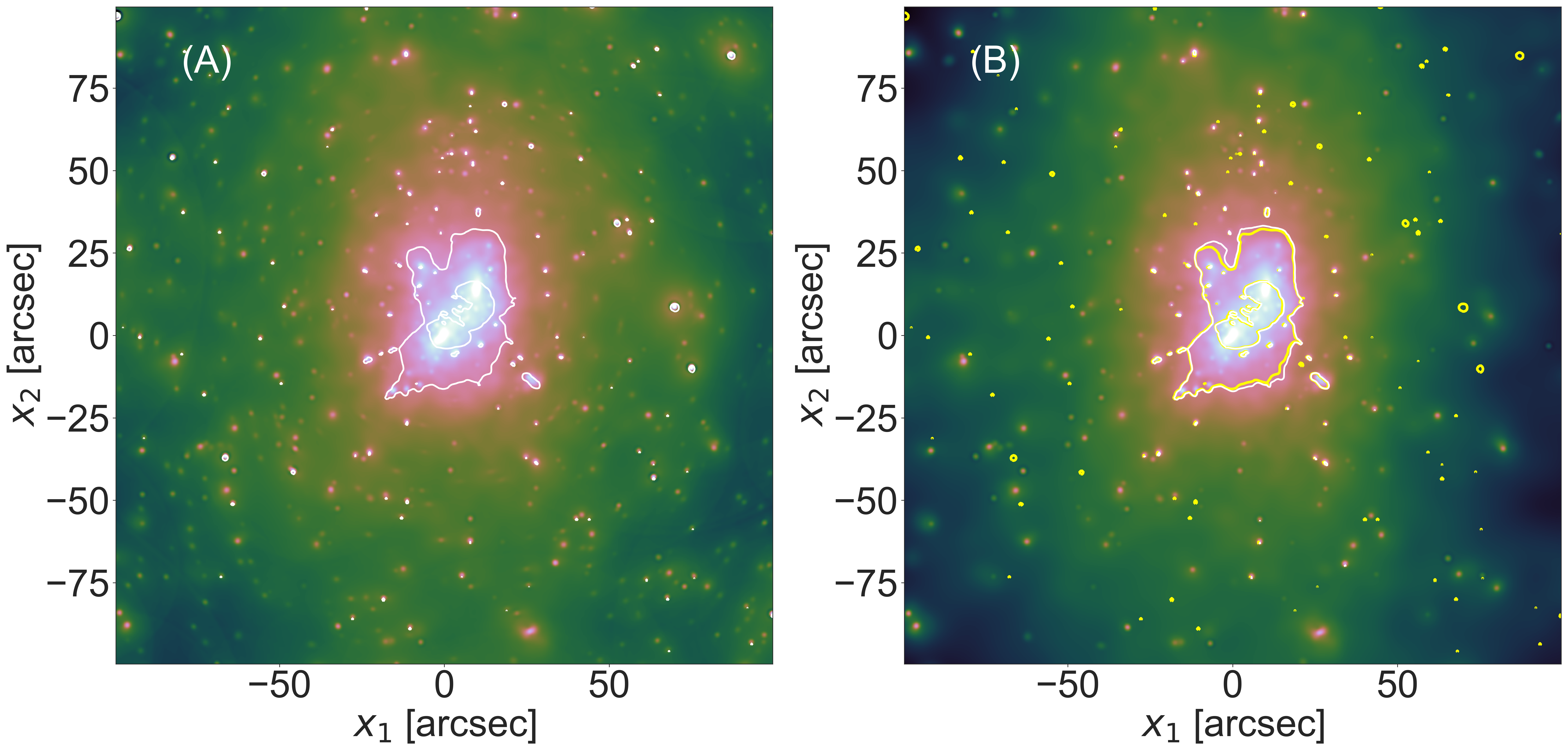}
\hfill
\caption{\label{fig:validation_tests2}  {\bf Convergence maps  with and without line-of-sight structure.}  Panels A and B show one projection of the cluster {\em Hera} \cite{2017MNRAS.472.3177M}, taken from the AGN sample. Panel A includes structures along the line-of-sight, while panel B does not. The critical curves for background sources at $z_s=7$ are shown in white. In panel B, we replicate the critical lines from panel A, now shown in yellow. This shows clearly that the critical lines of substructures within the cluster are not affected by the line-of-sight structure. In addition, all the secondary critical lines that appear after the inclusion of the line-of-sight structure do not correspond to any cluster mass substructure. The side length of each map is $200^{\prime\prime}$, corresponding to $0.88 h^{-1}\,$Mpc }.
\end{figure}

\end{document}